\documentclass[twocolumn,astrocommands]{aastex63}

\usepackage{multirow}
\usepackage{graphics}
\usepackage{rotating}
\usepackage{enumitem}
\usepackage{amsmath}
\usepackage{hyperref}
\usepackage{float}
\usepackage{tablefootnote}
\usepackage{footnote}
\usepackage{threeparttable}
\usepackage[mediumspace,mediumqspace,Grey,squaren]{SIunits}
\hypersetup{
			hyperfootnotes=true,
			colorlinks=true,
			linkcolor=blue,
			citecolor=blue,
			urlcolor=blue,
			breaklinks=false,
}
\newcommand{\dn}{$\rm D_{\rm n}4000$}

\newcommand{\ilum}{$L_{\rm i}/L^{*}$}
\newcommand{\HI}{\hbox{{\rm H}\kern 0.2em{\sc i}}}
\definecolor{LightCyan}{rgb}{0.88,1,1}

\shorttitle{Kinematic Approach to Environmental Quenching Since $z \sim 1$}
\shortauthors{Kim et al.}
\graphicspath{{./}{figures/}}
\begin{document}

\title{A Gradual Decline of Star Formation since Cluster In-fall:\\ New Kinematic Insights into Environmental Quenching at 0.3 $< z <$ 1.1}

\correspondingauthor{Keunho J. Kim}
\email{kim2k8@ucmail.uc.edu}
\author[0000-0001-6505-0293]{Keunho J. Kim}
\affiliation{Department of Physics, University of Cincinnati, Cincinnati, OH 45221, USA}

\author[0000-0003-1074-4807]{Matthew B. Bayliss}
\affiliation{Department of Physics, University of Cincinnati, Cincinnati, OH 45221, USA}

\author[0000-0003-1832-4137]{Allison G. Noble}
\affiliation{School of Earth and Space Exploration, Arizona State University, Tempe, AZ 85287, USA}
\affiliation{Beus Center for Cosmic Foundations, Arizona State University, Tempe, AZ 85287, USA}

\author{Gourav Khullar}
\affiliation{Department of Astronomy and Astrophysics, University of Chicago, 5640 South Ellis Avenue, Chicago, IL 60637, USA}
\affiliation{Kavli Institute for Cosmological Physics, University of Chicago, 5640 South Ellis Avenue, Chicago, IL 60637, USA}
\affiliation{Kavli Institute for Astrophysics $\&$ Space Research, Massachusetts Institute of Technology, 77 Massachusetts Ave., Cambridge, MA 02139, USA}

\author{Ethan Cronk}
\affiliation{Department of Physics, University of Cincinnati, Cincinnati, OH 45221, USA}

\author{Joshua Roberson}
\affiliation{Department of Physics, University of Cincinnati, Cincinnati, OH 45221, USA}

\author{Behzad Ansarinejad}
\affiliation{School of Physics, University of Melbourne, Parkville, VIC 3010, Australia}

\author{Lindsey E. Bleem}
\affiliation{Kavli Institute for Cosmological Physics, University of Chicago, 5640 South Ellis Avenue, Chicago, IL 60637, USA}
\affiliation{Argonne National Laboratory, High-Energy Physics Division, 9700 S. Cass Avenue, Argonne, IL 60439, USA}

\author[0000-0003-4175-571X]{Benjamin Floyd}
\affiliation{Department of Physics and Astronomy,
University of Missouri-Kansas City, Kansas City, MO 64110, USA}

\author{Sebastian Grandis}
\affiliation{Faculty of Physics, Ludwig-Maximilians-Universit{\"a}t, Scheinerstr. 1, 81679 Munich, Germany}

\author[0000-0003-3266-2001]{Guillaume Mahler}
\affiliation{Institute for Computational Cosmology, Durham University, South Road, Durham DH1 3LE, UK}
\affiliation{Centre for Extragalactic Astronomy, Durham University, South Road, Durham DH1 3LE, UK}

\author{Michael A. McDonald}
\affiliation{Kavli Institute for Astrophysics $\&$ Space Research, Massachusetts Institute of Technology, 77 Massachusetts Ave., Cambridge, MA 02139, USA}

\author[0000-0003-2226-9169]{Christian L. Reichardt}
\affiliation{School of Physics, University of Melbourne, Parkville, VIC 3010, Australia}

\author{Alexandro Saro}
\affiliation{Astronomy Unit, Department of Physics, University of Trieste, via Tiepolo 11, I-34131 Trieste, Italy}
\affiliation{IFPU - Institute for Fundamental Physics of the Universe, Via Beirut 2, 34014 Trieste, Italy}
\affiliation{INAF - Osservatorio Astronomico di Trieste, via G. B. Tiepolo 11, I-34143 Trieste, Italy}
\affiliation{INFN - National Institute for Nuclear Physics, Via Valerio 2, I-34127 Trieste, Italy}

\author[0000-0002-7559-0864]{Keren Sharon}
\affiliation{Department of Astronomy, University of Michigan, 1085 South University Avenue, Ann Arbor, MI 48109, USA}

\author{Taweewat Somboonpanyakul}
\affiliation{Kavli Institute for Particle Astrophysics \& Cosmology (KIPAC), 452 Lomita Mall, Stanford, CA 94305, USA}

\author{Veronica Strazzullo}
\affiliation{INAF - Osservatorio Astronomico di Trieste, via G. B. Tiepolo 11, I-34143 Trieste, Italy}
\affiliation{INAF – Osservatorio Astronomico di Brera, Via Brera 28, 20121 Milano, Via Bianchi 46, 23807 Merate, Italy}

%\collaboration{6}{(AAS Journals Data Editors)}
%==============================================================================



%=================================Abstract========================================
\begin{abstract}
The environments where galaxies reside crucially shape their star formation histories.
We investigate a large sample of 1626 cluster galaxies located within 105 galaxy clusters spanning a large range in redshift ($0.26 < z < 1.13)$.
The galaxy clusters are massive (M$_{500} \gtrsim 2\times10^{14}$M$_{\odot}$), and are uniformly selected from the SPT and ACT Sunyaev-Zel'dovich (SZ) surveys.
With spectra in-hand for thousands of cluster members, we use galaxies' position in projected phase space as a proxy for their in-fall times, which provides a more robust measurement of environment than quantities such as projected cluster-centric radius.
We find clear evidence for a gradual age increase of the galaxy's mean stellar populations ($\sim$ 0.71 $\pm$ 0.4 Gyr based on a 4000 $\rm{\AA}$ break, \dn ) with the time spent in the cluster environment.
This environmental quenching effect is found regardless of galaxy luminosity (faint or bright) and redshift (low-$z$ or high-$z$), although the exact stellar age of galaxies depends on both parameters at fixed environmental effects.
Such a systematic increase of \dn\ with in-fall proxy would suggest that galaxies that were accreted into hosts earlier were quenched earlier, due to longer exposure to environmental effects such as ram pressure stripping and starvation.
Compared to the typical dynamical time scales of $1-3$ Gyr of cluster galaxies, the relatively small age increase ($\sim$ 0.71 $\pm$ 0.4 Gyr) found in our sample galaxies seems to suggest that a slow environmental process such as starvation is the dominant quenching pathway.
Our results provide new insights into environmental quenching effects spanning a large range in cosmic time ($\sim 5.2$ Gyr, $z=0.26$--1.13) and demonstrate the power of using a kinematically-derived in-fall time proxy.
\end{abstract}
%==================================================================================

\keywords{galaxies: clusters --- galaxies: evolution --- galaxies: formation --- galaxies: star formation}

%=====Section 1=====
\section{Introduction---A Kinematic Method for Investigating Environmental Quenching at High-z}
\label{sec:introduction}
%\noindent 
The environment where galaxies reside is closely linked to the galaxies' characteristics, holding key information for their evolution histories.
In particular, cluster environments have shown remarkably tight relationships with the fundamental properties of galaxies such as morphology \citep{dres80,oh18}, star formation rate \citep{nobl13,muzz14,park09}, and gas components \citep[e.g.,][]{kenn04,chun09}.

Such close relationships between cluster environments and galaxy properties suggest that the extremely dense environment of clusters has indeed affected the residing galaxies through various effects such as hot intracluster medium (ICM) ram-pressure stripping \citep{gunn72,abad99,ebel14,bose21}, strangulation (starvation) \citep{lars80}, harassment \citep{moor96}, and tidal interactions \citep{byrd90}.

In particular, these environmental effects are known to efficiently suppress (a.k.a., quench) the star formation activity of cluster galaxies.
Indeed, the higher fraction of quiescent galaxies in clusters compared to field environments \citep{stra19,pint19,paul20} and systematically truncated {\HI} gas disks found in local cluster galaxies \citep[e.g., the Virgo cluster,][]{yoon17} suggest the cluster environmental ``quenching'' effects at play.

However, no clear consensus about the environmental effect has been established yet at high redshifts, contrary to the local Universe ($z < 0.2$) where a wealth of clusters have been observed and analyzed in detail \citep[e.g.,][for a review]{oh18,pasq19,smit19,upad21,moro21,bose21,cort21}.
The decreasing number of clusters and the reduced brightness with increasing redshift make it challenging to study high-redshift cluster populations.
While several studies have indeed investigated the environmental effects outside the local Universe \citep[e.g.,][]{stra13,pint19,kelk19,webb20,balo21,reev21,noor21,khul21}, only relatively small number of clusters at high redshift ($z \gtrsim 0.4$) have been studied in detail for environmental quenching effects with \textit{spectroscopically confirmed} member galaxies  \citep[e.g.,][]{muzz12,vaug20,tile20,balo21,math21}.
Furthermore, most studies at high redshift use the galaxies' projected distance from cluster center as a primary environmental measure for quenching effects in clusters.
While the projected clustercentric distance is a useful environmental parameter, it inevitably suffers from the contamination from interlopers that happen to lie within the cluster's projected radius, but in turn are unrelated to the host cluster.

Notably, by adopting an advanced environmental metric based on cluster galaxies' kinematic information, our study attempts to minimize the contamination from projected interlopers. 
Thus, our study enables us to investigate in detail how the star formation of galaxies changes over time since in-fall into the host cluster.
Specifically, we estimate galaxies' in-fall stages (ranging from recently accreted populations to early ancient in-fallers) by putting together their clustercentric distance \textit{and} peculiar velocity relative to the cluster center.

This ``phase-space'' (i.e., the diagram of the clustercentric distance versus the peculiar velocity normalized by the velocity dispersion of the cluster) is a powerful tool to study the detailed in-fall histories of cluster galaxies, since it considers not only the distance from the cluster center but also the ``kinematic'' velocity of in-falling galaxies.
Numerous cluster simulations, where galaxies' time-steps can be traced, have indeed demonstrated the clear separations of galaxies by different in-fall stages in phase-space \citep{maha11,oman13,muzz14,hain15,jaff15,oman16,rhee17,rhee20}.

Motivated by the simulation results, several studies utilized the \textit{projected} phase-space (i.e., the projected clustercentric radius and the line-of-sight peculiar velocity used instead) for actual cluster galaxies to estimate the galaxies' orbital stages over a wide range of redshifts, both local  \citep[e.g.,][]{voll01,maha11,hern14,bose14b,hain15,jaff18,gava18,shen20,loni21,reev22} and high-redshift clusters \citep[e.g.,][]{nobl13,muzz14,liu21}. In particular, \cite{nobl13} estimated the in-fall time of galaxies in a $z\sim0.9$ cluster. 
They found a systematic decrease of specific star formation rates (sSFR) of galaxies as a function of in-fall time proxy derived from the galaxies' location in the projected phase space, suggesting the projected phase space as a robust observational environmental measure for in-fall time.
Later studies in the local Universe \citep[$z < 0.2$ e.g.,][]{pasq19,smit19,samp21} support this by showing the similar trend of reduced sSFR of cluster galaxies based on the similar method using projected phase space.

The goal of our study is to provide a \textit{clearer} view of the environmental quenching effects than the projected distance alone, at high redshifts by crucially utilizing the advanced kinematic approach over a huge range in cosmic time ($\sim 5.2$ Gyr, $z=0.26$--1.13).
For that, we employ a large number of clusters uniformly selected by the Sunyaev-Zel'dovich (SZ) effect from South Pole Telescope (SPT \citealt{blee15}) and Atacama Cosmology Telescope (ACT \citealt{marr11}) cluster surveys.
We obtain the photometric and spectroscopic information of cluster galaxies from the optical follow-up observations of the SZ clusters \citep[][and references therein]{ruel14,blee15,bayl16}.

Section \ref{sec:Data Sets} describes the observational data sets.
Sample selection procedure and data analysis are described in Section \ref{sec:Sample Selection and Data Analysis}.
We present and discuss our results in Section \ref{sec:results and discussion}.
We summarize our conclusions with final remarks in Section \ref{sec:Summary and conclusions}. 
We adopt the $\Lambda$CDM cosmology of ($H_{0}$, $\Omega_{m}$, $\Omega_{\Lambda}$) = (70 $\rm{kms^{-1}}$ $\rm{Mpc^{-1}}$, 0.3, 0.7) throughout the paper.

%===============================2. Sample Selection and Data Analysis================================
\section{The Observational Data Sets}
\label{sec:Data Sets}
We use observational data sets drawn from the 2500 deg$^{2}$ SPT-SZ \citep{vand10,reic13,ruel14,blee15,bayl16,bayl17} and ACT-SZ \citep{marr11,hass13,sifo13} cluster surveys and the associated optical/near-IR follow-up observations.
Due to the redshift-independence of the SZ effect (SZE), the clusters identified by these surveys are nearly mass-limited with a mass threshold of $\gtrsim 3 \times 10^{14} M_{\odot}$ at all redshifts (see e.g., Figure 6 of \citealt{blee15}).
This uniform cluster selection function through the SZE, combined with the uniform selection of the galaxy samples, allows us to study environmental effects on galaxy properties in a uniform way over a wide range of redshift (0.26 $< z <$ 1.13).
In this section, we describe details about the cluster surveys (Section \ref{subsec:Clusters from SPT and ACT}), galaxy photometric $i$-band luminosity (Section \ref{subsec:Galaxy i band Luminosity}), and the spectral 4000 $\rm{\AA}$ break measurements (Section \ref{subsec:Spectral 4000 Break}).

%===============================2.1 SPT and ACT surveys via SZ effect================================
\subsection{Clusters Uniformly Selected by the Sunyaev-Zel'dovich Effect from SPT and ACT at 0.26 $< z <$ 1.13}
\label{subsec:Clusters from SPT and ACT}
Our sample of clusters is a sub-sample of the clusters identified in the 2500 deg$^{2}$ SPT-SZ \citep{reic13,blee15} and ACT-SZ surveys \citep{marr11,hass13}.
We refer the reader to the aforementioned publications for full descriptions of these surveys. 
In short, these cluster candidates were first detected through their SZ signal (i.e., the SZ signal-to-noise threshold $\xi \gtrsim$ 4.5), and subsequently followed up with optical/near-IR imaging to confirm the clusters associated with the SZ signal.
About 500 clusters (415 and 68 from the SPT and ACT, respectively with some overlapping clusters between the surveys) have been discovered in the surveys at redshifts $z = 0.1$--1.5.
The cluster masses---$M_{\rm 500}$ the mass measured within the radius $r_{\rm 500}$ at which the mean density of the cluster is 500 times the critical density at the cluster redshift---are $\gtrsim 3 \times 10^{14} M_{\odot}$, nearly independent of redshift due to the selection function based on the SZ effect.

From the surveys, we adopt information about the cluster center, redshift ($z_{\rm cl}$), and mass ($M_{\rm 500}$) of our 105 sample clusters at $0.26 < z_{\rm cl} < 1.13$ (see Section \ref{subsec:Sample Selection} for details about sample selection).
Specifically, 99 clusters from SPT and 9 from ACT satisfy these criteria,  with three clusters appearing in both catalogs. The result is a total sample of 105 SZ-selected galaxy clusters. 
For the three clusters in both catalogs, we use the SZ information from SPT alone (there are no published scaling relations for a combination of SPT and ACT SZ data, but see \citealt{hilt18} for the discussion of ACT and SPT mass consistency).
In particular, the cluster mass ($M_{\rm 500}$) is derived from the SZ signal--cluster mass scaling relation (for 99 SPT clusters measured from \citealt{reic13,blee15} and 6 ACT clusters measured from \citealt{hass13}).
For the SPT clusters, the following scaling relation is used:
%;============Equation 1 ==========
\begin{eqnarray}
\zeta = A_{\rm{SZ}} \left( \frac{M_{\rm 500}}{3 \times 10^{14} M_{\odot}h^{-1}} \right)^{B_{\rm SZ}} \left( \frac{H(z)}{H(0.6)} \right)^{C_{\rm SZ}} \ ,
\label{Eq1}
\end{eqnarray}
where $A_{\rm{SZ}}$ is the normalization factor, $B_{\rm{SZ}}$ the slope, and $C_{\rm{SZ}}$ the redshift evolution term associated with the Hubble parameter $H(z)$. $\zeta$ is the ``unbiased SZ significance\footnote{See Appendix B in \cite{vand10} for details.}'' associated with the SZ signal-to-noise threshold $\xi$ as follows:
\begin{eqnarray}
\zeta = \sqrt{\langle \xi \rangle^{2} - 3}
\label{Eq2}
\end{eqnarray}
The values of $A_{\rm{SZ}}$, $B_{\rm{SZ}}$, and $C_{\rm{SZ}}$ are 4.14, 1.44, and 0.59, respectively as determined in \cite{reic13}.
For the 6 ACT clusters, we adopt the SZ signal based-cluster mass that is calibrated with the cluster physics model of \cite{bode12} in \cite{hass13} (see Section 3.4 and Table 10 of \cite{hass13} for more details).

Further details on the cluster surveys such as cluster identification and cluster mass estimates can be found in \cite{blee15} and \cite{hass13} for SPT and ACT clusters, respectively.

%===============================Section 2.2 for galaxy i-band luminosity===============================
\subsection{Galaxy $i$-band Luminosity Relative to the Characteristic Luminosity ($L_{\rm i}/L^{*}$) for Photometric Brightness}
\label{subsec:Galaxy i band Luminosity}
We obtain the $i$-band luminosity ($L_{\rm i}$) of our sample galaxies from the optical follow-up observations of the SPT cluster surveys \citep[][and references therein]{blee15,blee20} that are conducted to confirm the clusters associated with the SZ signal.
Several different telescopes are used for the follow-up observations (see Table 2 of \citealt{blee15} and for more recent observations with the PISCO instrument \citep{stal14}, see Section 4 of \citealt{blee20}), including Blanco/MOSAIC-II; Magellan/Baade IMACS f-2; Magellan/Clay LDSS3, Megacam, and PISCO; Swope/SITe3; MPG/ESO WFI; New Technology Telescope/EFOSC2.

While the aperture size of the telescopes is not the same (1-6.5 m), we note that the follow-up campaign is optimally designed to observe clusters with uniformly sufficient depth at both low ($z\lesssim 0.3$) and high ($z\gtrsim 0.75$) redshifts.
Typically, the observations are required to detect 0.4$L^{*}$ galaxies with $5\sigma$ depth, where $L^{*}$ is the characteristic luminosity of galaxies at a given redshift.
The majority of the photometry use the SDSS $i$-band filter.
For clusters observed with the older Johnson–Cousins photometric $BVRI$ system, the filter transformation of the $I$-band into the SDSS $i$-band is applied in \cite{bayl16} (see their Section 5.2).
As a result, all of our sample galaxies' $L_{\rm i}$ is measured with respect to the SDSS $i$-band filter.

We use $i$-band luminosity ($L_{\rm i}$) scaled by $L^{*}$, \ilum, as our sample galaxies's photometric brightness.
We adopt the $i$-band $L^{*}$ values computed in the SPT-SZ surveys \citep{high10,blee15}.
Specifically, the $L^{*}$ is derived using the stellar population synthesis model of \cite{bruz03}, and
the stellar population modeling assumes a $k$-corrected passively evolving, instantaneous-burst stellar population with a formation redshift of $z=3$.
The Salpeter initial mass function \citep{salp55} and the Padova 1994 stellar evolutionary tracks \citep{fago94} are used.
Metallicities are selected based on analytic fits to the Red-sequence Cluster Survey 2 data \citep{gilb11}.
Interpolation using cubic spines is applied to generate $L^{*}$ values at redshifts where the stellar synthesis model did not directly compute $L^{*}$.
The models are further calibrated to the actual SPT spectroscopic sub-sample.
A comparison shows that the computed $L^{*}$ for our sample cluster galaxies is in good agreement with that of the maxBCG cluster sample of \cite{ryko12}.
Specifically, the $L^{*}$ values between the two studies are consistently better than $\lesssim 5 \%$ over the overlapping redshift range at $z= 0.05$--0.35.
As we will describe in Section \ref{subsec:Sample Selection}, we use galaxies brighter than 0.35 \ilum as our sample galaxies across all redshift of interest ($0.26 < z < 1.13$).

%----------Figure 1-------
\begin{figure*}[t!]
\centering
\includegraphics[width=1\textwidth]{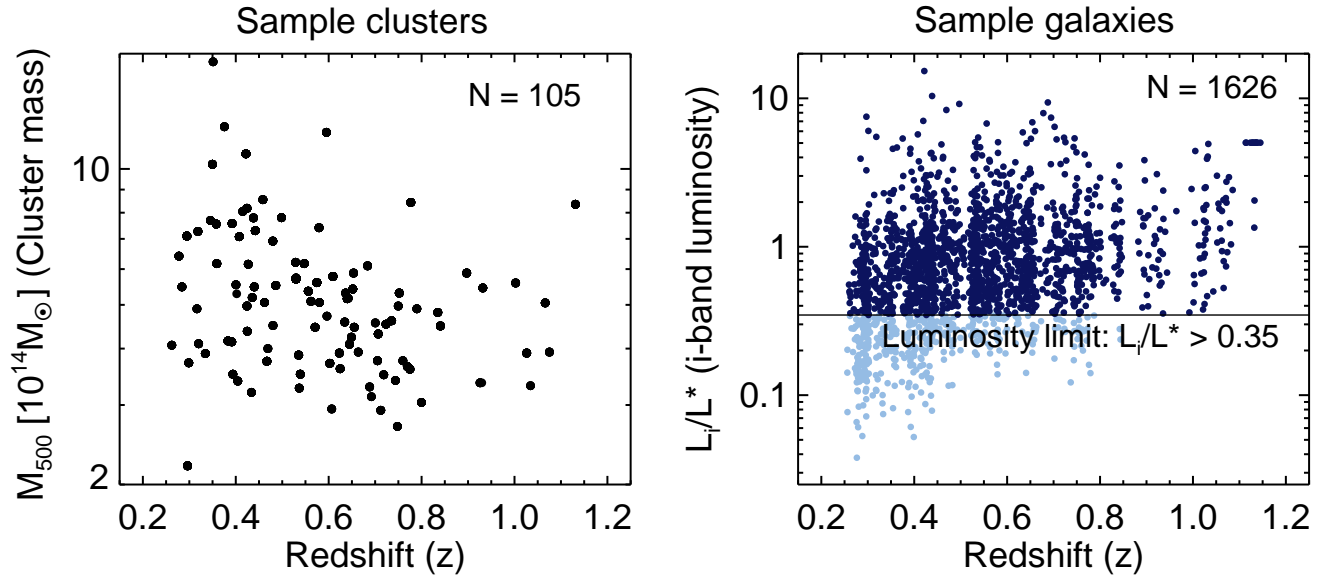}
\caption{\textbf{Left:} the sample cluster mass distribution with redshift. Due to the uniform selection function based on the redshift-independent SZ effect, our sample clusters are nearly mass-limited at all redshifts, forming a relatively narrow mass range with a median (standard deviation) mass of $4.97 \ (\pm \ 2.24) \times 10^{14} M_{\odot}$ (Section \ref{subsec:Clusters from SPT and ACT}).
\textbf{Right:} the $i$-band luminosity (\ilum , relatively to the characteristic luminosity) distribution of sample cluster galaxies with redshift.
The dark blue points are the sample galaxies used after applying the luminosity cut of \ilum \ $> 0.35$, while the light blue points are the ones excluded.
The number of sample clusters and cluster galaxies considered in this study are marked on top right in the left and right panels, respectively.
See also Table \ref{tab1} and Section \ref{subsec:Sample Selection} for details on the sample selection.}
\label{fig1}
\end{figure*}

%;--------------------Section 2.3 for galaxy 4000 A break-----------
\subsection{The Age-Sensitive 4000 $\rm{\AA}$ Break $\rm{(}$\dn $\rm{)}$ for the Mean Stellar Age of Galaxies}
\label{subsec:Spectral 4000 Break}
We use the 4000 $\rm{\AA}$ break strength (\dn) as a proxy for the luminosity-weighted mean stellar age of galaxies.
\dn\ is a spectral feature that arises from the accumulation of the metal absorption lines of stars making a ``break'' between the blue- and red-side continua around 4000 $\rm{\AA}$ in rest frame wavelength \citep{bruz83,hami85,balo99}.
Because the metal absorption line strength is sensitive to stellar type (i.e., age given surface gravity and temperature), \dn\ is sensitive to the luminosity-weighted mean stellar age of galaxies.
Thus, \dn\ has been widely adopted as a useful stellar age indicator of galaxies \citep[][and reference therein]{kauf03,hern13,hain17,kim18}.

\dn\ is defined as the average flux density ($\langle F \rangle$) ratio between the blue-side ($\langle F^{B} \rangle$) and the red-side ($\langle F^{R} \rangle$) continua centered at 4000 $\rm{\AA}$, such that

%------------Dn4000 definition------------------------------------
\begin{eqnarray}
\langle F \rangle =  \frac{\int_{\lambda_1}^{\lambda_2} F_{\nu} \ d{\lambda}}{\int_{\lambda_1}^{\lambda_2}  d{\lambda}},
\label{Eq3}
\end{eqnarray}
and thus, \dn\ is expressed as:
\begin{eqnarray}
 {\rm D}_{\rm n}4000 = \frac{\langle F^{R} \rangle}{\langle F^{B} \rangle} = \frac{(\lambda_2^B-\lambda_1^B) \int_{\lambda_1^R}^{\lambda_2^R} F_{\nu} \ d{\lambda}}{(\lambda_2^R-\lambda_1^R) \int_{\lambda_1^B}^{\lambda_2^B} F_{\nu} \ d{\lambda}},
\label{Eq4}
\end{eqnarray}
where ($\lambda_{1}^{B}$, $\lambda_{2}^{B}$, $\lambda_{1}^{R}$, $\lambda_{2}^{R}$,) = (3850, 3950, 4000, 4100) $\rm{\AA}$.
\dn\ increases with galaxy age such that galaxies with young stellar populations ($\lesssim 1$ Gyr) show \dn\ $\lesssim 1.5$, while galaxies with old stars ($\gtrsim 1$ Gyr) show \dn\ $\gtrsim 1.5$ based on simple stellar population modelings \citep[see e.g., Figure 2 of ][]{kauf03}.
%---------------------------------------------------------

We obtain the \dn\ of our sample galaxies measured from the optical follow-up spectroscopic observations of SPT-SZ clusters \citep[][and references therein]{ruel14,bayl16,bayl17} and ACT-SZ clusters \citep{sifo13}.
Most clusters are observed with the Gemini Multi-Object Spectrograph (GMOS) on Gemini South, IMACS on Magellan/Baade, or FORS2 on VLT.
The observations are primarily designed to measure spectroscopic redshifts of galaxies for accurate cluster memberships through the cluster velocity dispersion ($\sigma_{\rm cl}$).

The specific observing strategy (e.g., target selection, multislit mask design, and the choice of grating and filter) is set up to uniformly observe a large number ($\gtrsim 100$) of clusters. The spectroscopic integration times for individual masks were computed to ensure that the signal-to-noise ratio is $\gtrsim 3$ per spectral pixel in the continuum around \dn, based on a model passive galaxy spectrum with a brightness equal to 0.4 \ilum\ at cluster redshift. 
Note that we quite consistently sample this luminosity range (see the right panel of Figure \ref{fig1}).

The target selection is based on the color-magnitude diagram of galaxies observed in the field of view of a given cluster.
The highest priority is given to cluster galaxies identified by the red sequence, regardless of luminosity down to $\simeq 0.4$ \ilum.
We first identified the red sequence as an over-density in color, and then fit a tilted red sequence to the data in color magnitude. 
Likely red sequence galaxies were those galaxies within $\pm$ 0.15 magnitudes (in color) of the best-fit red sequence.
The choice of $\pm$ 0.15 magnitudes corresponds to $\pm$ 2.5--3 $\sigma_{\rm RS}$, where $\sigma_{\rm RS}$ is the intrinsic width of the observed red sequence, which has been measured to be $\simeq$ 0.05 magnitudes for massive galaxy clusters \citep{henn17}.
The next highest priority is given to candidate ``blue-cloud'' galaxies, which are identified as those galaxies that are bluer than the red sequence (i.e., star-forming). We also only include galaxies that are fainter than the brightest red sequence member to filter out foreground galaxies. 

Obtaining spectroscopic follow-up of a complete sample of cluster galaxies in hundreds of distant galaxy clusters is not practical given the observational resource cost, and so it is not a realistic possibility to measure \dn\ for every cluster member galaxy in our cluster sample. This limitation prevents us from interpreting our results as an absolute measurement of the \dn\ of the complete sample of SPT cluster member galaxies. That said, we note that the slit placement strategy that produced our spectroscopic member sample was \textit{uniformly} applied to all spectroscopic observations, with no systematic evolution in the radial density of spectroscopic slits or the rest-frame magnitude limits. Because the slit placement strategy was applied uniformly, any systematic \dn\ difference \textit{within} our sample that scales with other parameters (e.g., galaxy luminosity, redshift, and environment) should be robust.
 
Understanding our spectroscopic galaxy sample selection is essential to interpreting our results, and so we also perform a direct comparison of our spectroscopic cluster member sample against a published photometric analysis that uses complete cluster galaxy populations. Specifically, we measure the fractions of passive (red sequence) and star-forming (blue cloud) cluster member galaxies in our final spectroscopic catalog in our two redshift bins. We measure passive (star-forming) fractions of $\sim$70 (30)\% and $\sim$66 (34)\% for low-redshift ($\langle z \rangle$ of 0.41) and high-redshift ($\langle z \rangle$ of 0.66) clusters, respectively. These fractions are consistent with the observed red/passive galaxy fraction measured for all SPT clusters using photometric data (\citealt{henn17}{, see their Figure 15}). The consistency between the passive and star-forming fractions in our spectroscopic catalog and published imaging analyses indicates that the sparse spectroscopic sampling does, on average, recover a representative population of cluster galaxies.

Typically, $N \lesssim 40$ galaxies per cluster are observed.
The observed spectra cover galaxy rest-frame wavelength $\simeq$ 3500-5150 $\rm{\AA}$, across redshifts $z\simeq 0.25$--1.1.
The observations have similar spectral resolution $d\rm{\AA}$ $\simeq$ 5--10 $\rm{\AA}$, corresponding to $R$ (=$\lambda/d\lambda$) $\simeq$ 500-1200.
The observed spectroscopy is used to measure \dn\ by \cite{bayl16} using the same definition given in Equations \ref{Eq3} and \ref{Eq4}.

We refer the reader to the spectroscopic follow-ups of SPT-SZ clusters \citep{ruel14,bayl16,bayl17} for more details about the observing strategy and the \dn\ measurements.

%---------------Section 3 for Sample Selection and Data Analysis----
\section{Sample Selection and Data Analysis}
\label{sec:Sample Selection and Data Analysis}

%---------------Sub-Section 3.1 for Sample Selection----
\subsection{Sample Selection: A Uniform Set of 105 Clusters and 1626 Galaxies at 0.26 $< z <$ 1.13}
\label{subsec:Sample Selection}
We select our sample clusters by cross-matching the SZ cluster catalogue and the optical photometric and spectroscopic follow-up catalogues described in Section \ref{sec:Data Sets}.
From the cross-match, we obtain cluster mass ($M_{\rm 500}$), galaxy luminosity (\ilum), \dn, and cluster velocity dispersion ($\sigma_{\rm cl}$). 
There are 105 clusters and 4089 galaxies in our initial sample with a redshift ranging from 0.26 to 1.13.
Of the 105 clusters, 99 of them are SPT clusters and the remaining 6 clusters are ACT clusters.
The cluster mass ranges $3 \times 10^{14} M_{\odot} \lesssim M_{\rm 500} \lesssim 3 \times 10^{15} M_{\odot}$ with a median mass of $4.97 \times 10^{14} M_{\odot}$ and the associated 1$\sigma$ population spread of $\pm \ 2.24 \times 10^{14} M_{\odot}$.
The mass distribution of our sample clusters is shown in the left panel of Figure \ref{fig1}.

Sampling galaxies with reliable \dn\ measurements is crucial in our analysis.
We apply additional sample selection criteria based on physically reasonable \dn\ values from reliable measurements, so these cuts remove galaxies whose spectra suffer from data reduction artifacts (e.g., bright sky line residuals) as well as high noise.
Specifically, we select galaxies whose \dn\ satisfies: 0.7 $<$ \dn\ $<$ 2.3; $\sigma$(\dn)/\dn\ (relative error) $<$ 0.3, leaving 2709 galaxies.
The \dn\ range selected reasonably covers the observed \dn\ strengths across all types of galaxies, corresponding to from young ($\sim$ 10 Myr) to old ($\sim$ 15 Gyr) stellar ages based on simple stellar population modeling \citep[e.g.,][]{kauf03,hern13}.

We also apply the luminosity cut of \ilum\ $>$ 0.35 to the remaining sample galaxies to ensure a uniform lower limit of galaxy luminosity at all redshifts.
Specifically, we derive the luminosity cut by considering the 10th percentile (i.e., above which 90 $\%$ of the population lie) of the \ilum\ distribution of our high-redshift (0.53 $< z <$ 1.13) sub-samples.
This corresponds to 0.35 of \ilum , which is very similar to the typical depth ($5\sigma$ for 0.4$L^{*}$) of the photometric follow-ups of SPT-SZ clusters \citep{blee15}.
We thus remove galaxies with \ilum $< 0.35$, leaving 2125 galaxies.
The \ilum\ distribution of our sample galaxies with the luminosity cut (\ilum $>$ 0.35) applied is shown in the right panel of Figure \ref{fig1}.
The \ilum\ range of the sample galaxies roughly corresponds to $10.3 \lesssim {\rm{log}}(M_{star}/M_{\odot}) \lesssim 11.6$, given the typical characteristics stellar mass of $10.8 \simeq {\rm{log}}(M_{star}/M_{\odot})$ \citep[e.g.,][]{adam21} at similar redshifts and $I$-band mass-to-light ratio of $0.8-2.5$ \citep[e.g.,][]{mcga14}.

Lastly, of 2125 galaxies, 499 galaxies are further removed based on their projected clustercentric radius ($r_{\rm proj}/r_{\rm 500}$) and the normalized line-of-sight peculiar velocity ($\Delta \rm{v}/\sigma_{\rm cl}$, Equation \ref{Eq5}).
This criterion is applied to select cluster member galaxies with reasonable orbital stages (i.e., $|\Delta$v$|$/$\sigma_{\rm cl} < $ 3.5 and $\rm{r}_{proj}$/$\rm{r}_{500} < $ 3), as we will describe in detail in Section \ref{subsec:Infall Time Estimate}.

In total, our final sample throughout the paper is 105 clusters and 1626 galaxies at 0.26 $< z <$ 1.13.
On average, 15 galaxies per cluster are sampled.
The typical uncertainty in cluster mass in our cluster sample is $0.9 \times 10^{14} M_{\odot}$, and the typical uncertainty in the $i$-band photometry of cluster member galaxies is 0.06 magnitudes.
The sample selection procedure is summarized in Table \ref{tab1}.

%--------------------------Table 1; Sample Selection--------------------------------
\begin{table*}[ht]
\centering
\begin{threeparttable}
\caption{Summary of Sample Selection (Section \ref{subsec:Sample Selection})}
\begin{tabular}{ll}
\hline \hline
Criterion & Explanation / (Number of galaxies)\\
\hline
Cross-matching the cluster catalogue and & To obtain cluster mass ($M_{\rm 500}$), redshift, velocity dispersion ($\sigma_{\rm cl}$), \\ 
optical photometric and spectroscopic follow-up & galaxy $i$-band luminosity (\ilum ), spectroscopic redshift, \\
catalogues\tnote{a} & and \dn\ / (4,089) \\
 \hline
0.7 $<$ $\rm D_{\rm n}$4000 $<$ 2.3 $\&$ & To remove the poor or missing measurements of \dn\ due to \\
$\sigma$($\rm D_{\rm n}$4000)/$\rm D_{\rm n}$4000 $<$ 0.3 & wavelength coverage of observed spectra / (2,709) \\
\hline
$0.26 < z < 1.13$ & Redshift range divided into low-$z$ ($0.26 < z < 0.53$) \\
& and high-$z$ ($0.53 < z < 1.13$) bins with 46 and 59 clusters, respectively\\
\hline
\ilum \ $> 0.35$\tnote{b} & Galaxy luminosity cut to ensure a uniform lower limit of
galaxy luminosity\\
& at all redshifts / (2,125)\\
\hline
Cluster member galaxies: & Membership based on the normalized line-of-sight peculiar velocity ($|\Delta$v$|$/$\sigma_{\rm cl}$)\\ 
$|\Delta$v$|$/$\sigma_{\rm cl} < $ 3.5 and $\rm{r}_{proj}$/$\rm{r}_{500} < $ 3 & and the projected clustercentric radius ($\rm{r}_{proj}$/$\rm{r}_{500}$) / \\
& (1626: 802 and 824 in low-$z$ and high-$z$ bins, respectively) \\
\hline
Total number of sample & 105 clusters and 1626 member galaxies\\
\hline \hline \\
\label{tab1}
\end{tabular}
{\small
\begin{tablenotes}
\item[a] The catalogues for SPT-SZ clusters \citep{reic13,blee15} and ACT-SZ clusters \citep{hass13}, and the optical photometric \citep{high10,blee15} and spectroscopic \citep{sifo13,ruel14,bayl16,bayl17} follow-ups (Section \ref{sec:Data Sets}).
\item[b] Luminosity cut of \ilum\ $>$ 0.35 estimated from the 90th percentile of the high-$z$ sub-samples (Section 3.1 and the right panel of Figure \ref{fig1}).
\end{tablenotes}
}
\end{threeparttable}
\end{table*}
%--------------------------

%;=============Section 3.2 in-fall time estimate=================
\subsection{Kinematically Estimating the In-fall Time of Galaxies from Projected Phase Space}
\label{subsec:Infall Time Estimate}
\begin{figure*}[ht!]
\centering
\includegraphics[width=0.9\textwidth]{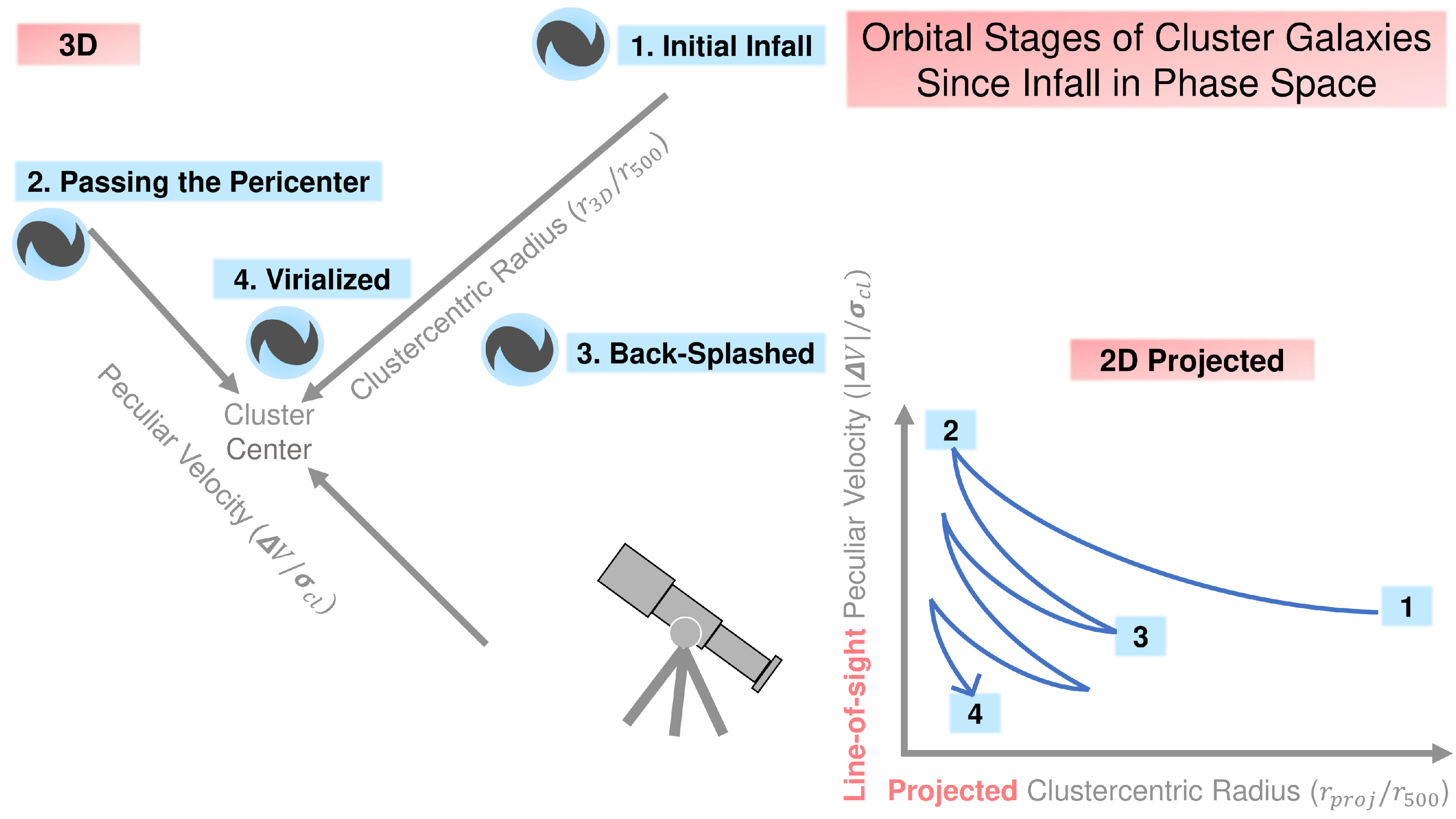}
\caption{
Schematic diagram showing the orbit of an in-falling galaxy into a cluster. 
This is illustrated in 3D (left) and 2D projected (right) phase space (i.e., peculiar velocity vs. clustercentric radius).
Moving from 1 to 4 stages, (1) an in-falling galaxy plunges into the gravitational potential of the cluster, (2) passing the pericenter, (3) back-splashed, and eventually (4) virialized.
This simplified orbit of an in-falling galaxy is qualitatively consistent with that of an in-falling galaxy traced in cluster simulations  \cite[e.g., see Figure 4 of][]{jaff15}.
This figure illustrates how a galaxy orbit is statistically traced in a phase space diagram during in-fall (see also Figure \ref{fig2} for the projected phase space of our sample galaxies).
}
\label{fig2_sch}
\end{figure*}

We estimate the in-fall time of galaxies based on their location in the phase space, similar to other studies \citep{nobl13,nobl16,pasq19}.
The phase space is a diagram where the in-fall stage of cluster galaxies can be kinematically estimated (Figure \ref{fig2_sch}).
It uses the clustercentric distance and the peculiar velocity normalized by cluster velocity dispersion on its axes.
For the \textit{projected} phase-space, the projected clustercentric distance ($r_{\rm proj}/r_{\rm 500}$) and the line-of-sight peculiar velocity normalized by the cluster velocity dispersion ($\Delta$v/$\sigma_{\rm cl}$) are employed.

Indeed, cluster simulations where the in-fall stages of galaxies can be traced have demonstrated that galaxies distinctively occupy different locations of phase space depending on their in-fall time \citep{mamo04,gill05,maha11,hain12,oman13,rhee17}.
Typically, galaxies that were accreted early show a wide range of peculiar velocities ($\Delta \rm{v}/\sigma_{\rm cl}$) at small clustercentric distances ($r/r_{\rm 500}$), but only populate small peculiar velocities over larger clustercentric distance.
In contrast, galaxies that were accreted recently are distributed over ranges of clustercentric distance and peculiar velocity, but preferentially distributed following trumpet-like profiles which can be described by lines of constant ($r/r_{\rm 500}$) $\times$ ($|\Delta \rm{v}|/\sigma_{\rm cl}$) \citep[see e.g., Figure 3 of ][]{hain12}.

\begin{figure*}[ht!]
\centering
\includegraphics[width=0.7\textwidth]{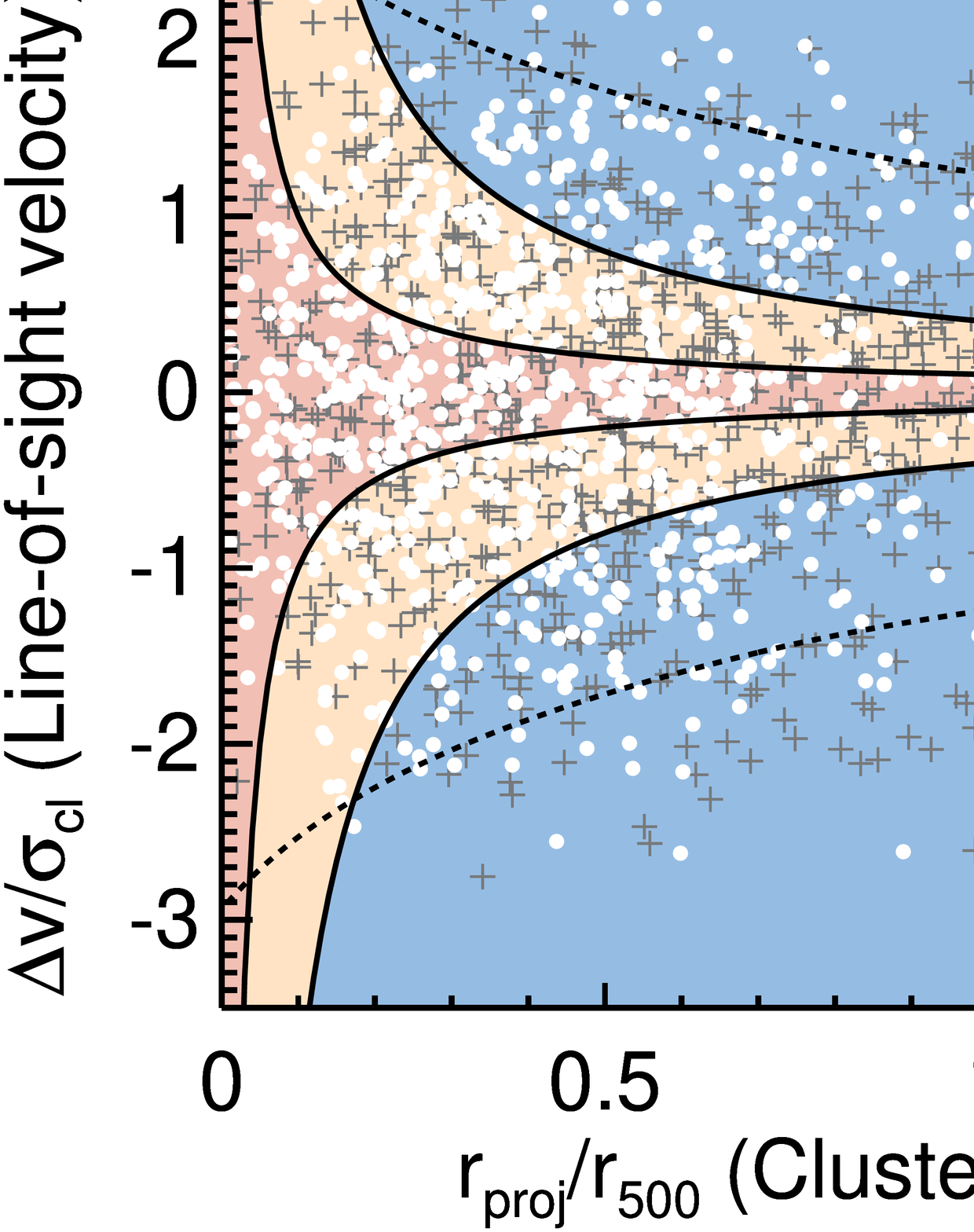}
\caption{The projected phase space diagram of galaxies stacked from 105 clusters at $0.26 < z < 1.13$ (i.e., the normalized line-of-sight peculiar velocity vs. projected clustercentric radius), which shows a variety of accretion stages of cluster galaxies since in-fall. 
The white circles and gray crosses indicate the low-redshift ($0.26 < z < 0.53$) and high-redshift ($0.53 < z < 1.13$) sub-sample galaxies, respectively.
The phase space is color coded by different in-fall time zones based on the caustic lines (i.e., solid lines, Section \ref{subsec:Infall Time Estimate}): Early in-fall (red): ($r_{\rm proj}/r_{\rm 500}$) $\times$ ($|\Delta \rm{v}|/\sigma_{\rm cl}$) $< 0.1$; Intermediate in-fall (orange): $0.1 < $ ($r_{\rm proj}/r_{\rm 500}$) $\times$ ($|\Delta \rm{v}|/\sigma_{\rm cl}$) $< 0.4$; Recent in-fall (blue): ($r_{\rm proj}/r_{\rm 500}$) $\times$ ($|\Delta \rm{v}|/\sigma_{\rm cl}$) $ > 0.4$.
The dashed line is the escape velocity curve of a cluster assuming an NFW potential \citep{nava96} with the typical cluster mass ($M_{\rm 500} =  4.97 \times 10^{14} M_{\odot}$) and velocity dispersion ($\sigma_{\rm {cl}} = 1052 \ \rm{km} \  \rm{s}^{-1}$) of our sample clusters.
Comparison of galaxies' distribution with the escape velocity curve suggests that galaxies within Early and Intermediate in-fall zones are likely gravitationally bound to the cluster, while galaxies within the Recent in-fall zone are becoming gravitationally bound.
This figure shows the overall accretion stages of cluster galaxies over a wide range of redshift and illustrates how our analysis adopts galaxies' location in the projected phase space to statistically infer the mean in-fall time of galaxies based on cluster simulations (Section \ref{subsec:Infall Time Estimate}).
}
\label{fig2}
\end{figure*}

Motivated by these phase space trends with in-fall time in simulations, \cite{nobl13} \textit{observationally} utilized the caustic lines---the trumpet-shaped lines satisfying ($r_{\rm proj}/r_{\rm 500}$) $\times$ ($|\Delta \rm{v}|/\sigma_{\rm cl}$) = constant (see also Figure \ref{fig2})---of the projected phase space to divide galaxies into different in-fall time stages for a $z \sim 0.9$ cluster.
Their analysis and later studies \citep{nobl16,pasq19} have demonstrated the utility of the caustic lines as a useful in-fall time proxy for cluster galaxies. 
Especially, a comparison with cluster simulations in \cite{pasq19} shows that galaxies separated by the caustic lines in the projected phase space indeed show systematically different mean in-fall time, as described below.

%----------Figure 2---------

%----------Figure 3---------

We adopt the same definition of in-fall time zones based on the location of phase space as in \cite{nobl13} which is as follows:\footnote{While the original definition of \cite{nobl13} is based on $r_{\rm 200}$ for a cluster radius, we note that using $r_{\rm 500}$ instead makes only a 0.15 dex shift in our in-fall time proxy (i.e., log[($r_{\rm proj}/r_{\rm 500}$) $\times$ ($|\Delta \rm{v}|/\sigma_{\rm cl}$]) and thus does not change our results qualitatively given $r_{\rm 500} \simeq 0.7r_{\rm 200}$ \citep[e.g.,][]{etto09}.}

\begin{enumerate}
    \item[(i)] Early in-fall: ($r_{\rm proj}/r_{\rm 500}$) $\times$ ($|\Delta \rm{v}|/\sigma_{\rm cl}$) $< 0.1$
    \item[(ii)] Intermediate in-fall: $0.1 < $ ($r_{\rm proj}/r_{\rm 500}$) $\times$ ($|\Delta \rm{v}|/\sigma_{\rm cl}$) $< 0.4$
    \item[(iii)] Recent in-fall: ($r_{\rm proj}/r_{\rm 500}$) $\times$ ($|\Delta \rm{v}|/\sigma_{\rm cl}$) $ > 0.4$ \\
\end{enumerate}
{\setlength{\parindent}{0pt} As the names of each in-fall zone indicate, the `Early' in-fall zone primarily consists of virialized galaxies that were accreted early.}
The `Intermediate' in-fall zone contains a mix of galaxy populations with intermediate in-fallers as well as a backsplash population which have passed the first cluster pericenter and are currently outbound.
The `Recent' in-fall zone is mainly populated by galaxies that were recently accreted.
The mean in-fall time of galaxies is typically $2.7^{+1.4}_{-1.7}$ Gyr, $4.5^{+1.9}_{-2.2}$ Gyr, and $5.1^{+1.8}_{-2.1}$ Gyr for Recent, Intermediate, and Early in-fall zones, respectively.
This is based on the comparison with the cluster simulation at $z=0$ in \cite{pasq19} (However, we also note that there is scatter in the mean in-fall time derived from the projected phase space due to projection effects as discussed in Section \ref{subsec:projection effects}).

For the projected phase space of sample clusters, $\Delta$v and $\sigma_{\rm cl}$ are based on the spectroscopic observations of SPT clusters (Section \ref{subsec:Spectral 4000 Break}).
In particular, the line-of-sight peculiar velocity ($\Delta$v) is expressed as follows:
%====================Peculiar line-of-sight velocity, Equation 5=========================
\begin{eqnarray}
\Delta \rm{v} = \textit{c} \times \left( \frac{\textit{z}_{gl}-\textit{z}_{cl}}{1+\textit{z}_{cl}} \right),
\label{Eq5}
\end{eqnarray}
%======================================================
where $z_{\rm gl}$, $z_{\rm cl}$, and $c$ are the redshifts of a galaxy and a cluster, and the speed of light, respectively.
The $r_{\rm proj}$ is derived by measuring the angular separation between the cluster center and a galaxy's projected location in the units of Right ascension (RA) and Declination (Dec).
To obtain the clustercentric radius ($\rm{r}_{proj}$/$\rm{r}_{500}$), we derive the $r_{\rm 500}$ of our sample clusters using the cluster mass (Section \ref{subsec:Clusters from SPT and ACT}) and assuming a spherical mass density profile:
%====================Equation 6, r500 and M500 =========================
\begin{eqnarray}
M_{\rm 500} = \frac{4\pi}{3}r_{\rm 500}^{3}500\rho_{\rm crit} ,
\label{Eq6}
\end{eqnarray}
%======================================================
where $\rho_{\rm crit}$ is the critical density of the Universe at cluster redshift.

We select galaxies with $|\Delta$v$|$/$\sigma_{\rm cl} < $ 3.5 and $\rm{r}_{proj}$/$\rm{r}_{500} < $ 3 as in-falling cluster galaxies as stated in Section \ref{subsec:Sample Selection}.
Figure \ref{fig2} shows the projected phase space of sample galaxies.
Galaxies mostly populate regions within $|\Delta$v$|$/$\sigma_{\rm cl} \lesssim 2$ and $\rm{r}_{proj}$/$\rm{r}_{500} \lesssim 1.5$.
These regions are closely matched with the inner region of the escape velocity radial profile (the dashed line) of the median mass ($4.97 \times 10^{14} M_{\odot}$) of our sample clusters.
This suggests that the bulk of in-falling galaxies are gravitationally bound by cluster potential wells.
To derive the escape velocity profile, we assume an NFW dark matter halo density profile \citep{nava96} and a cluster concentration parameter of 4.

Early and Intermediate in-fall zones are all located within the inner region of the escape velocity profile, while the Recent in-fall zone spans across the escape velocity profile.
The comparison of in-fall zones with the escape velocity profile makes intuitive sense in that galaxies that have in-fallen at earlier times are expected to have virialized earlier than those that have in-fallen more recently.
Our low-$z$ ($0.26 < z < 0.53$) and high-$z$ ($0.53 < z < 1.13$) sub-samples (i.e., gray and white points, respectively in Figure \ref{fig2}) show similar phase space distributions.

%===================4. Results and Discussion==========================
\section{Results and Discussion}
\label{sec:results and discussion}
In this section, we look at how the mean age of galaxies (using \dn) statistically varies with their in-fall time proxy to investigate environmental quenching effects since galaxies enter into clusters.
In Section \ref{subsec:entire quenching trend}, we show the trends of \dn\ with in-fall time proxy across  redshifts $z= 0.26$--1.13.
In Section \ref{subsec:luminosity dependence on the quenching}, we investigate the galaxy luminosity (using \ilum) dependence of the environmental quenching by dividing galaxies into `faint' (sub-$L^{*}$, \ilum\ $< 1$) and `bright' (super-$L^{*}$, \ilum\ $> 1$) galaxies.
In Section \ref{subsec:redshift and luminosity dependence}, we simultaneously control for galaxy luminosity and redshift dependences of the environmental quenching by dividing galaxies into redshift and luminosity bins.
We discuss the potential projection effects in our projected phase space analysis in Section \ref{subsec:projection effects}.

%==================Section 4.1 =============================
\subsection{Quenching Since In-fall: A Continuous \dn\ Increase of Galaxies from Recent to Early In-fall}
\label{subsec:entire quenching trend}

%----------Figure 3----------Environmental quenching trend only (i.e., no mass quenching)
\begin{figure*}[!ht]
\centering
\includegraphics[width=0.6\textwidth]{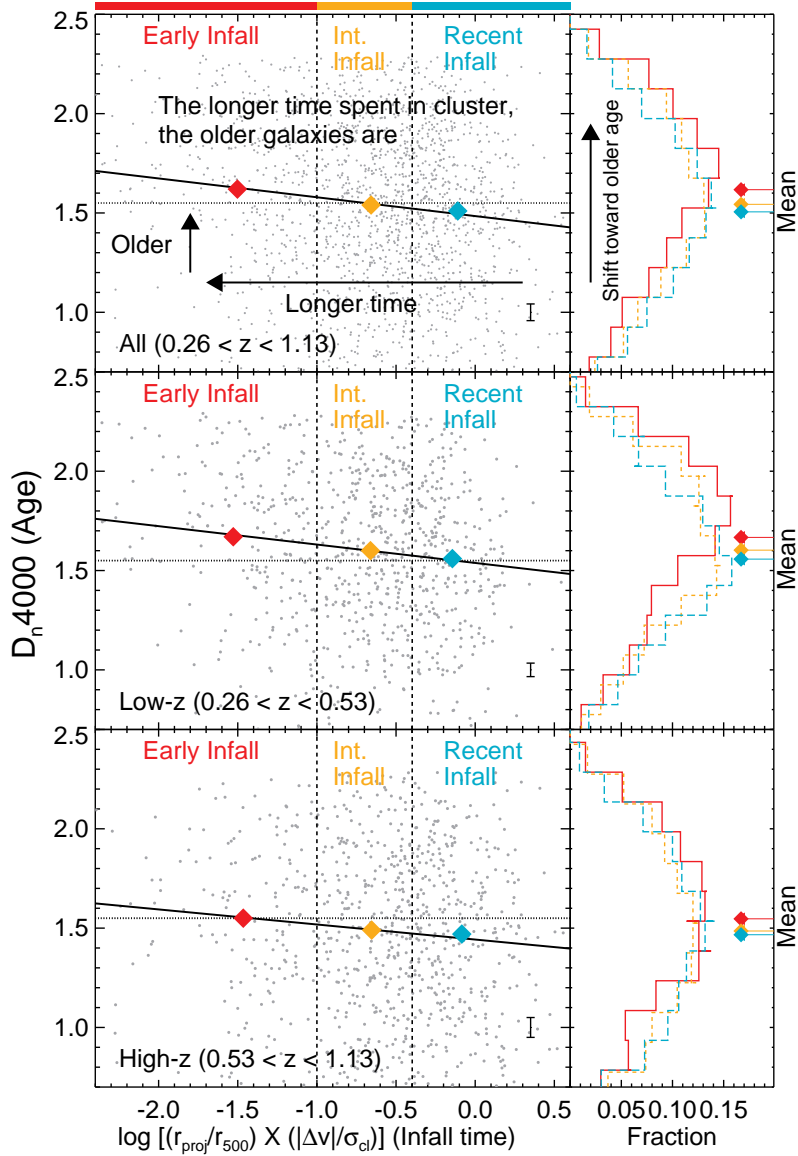}
\caption{Diagram showing that galaxies that have in-fallen at earlier times are quenched earlier across all in our sample redshifts ($z = 0.26-$1.13).
The 4000 $\rm{\AA}$ break (\dn ) and the position in the projected phase space (($r_{\rm proj}/r_{\rm 500}$) $\times$ ($\Delta \rm{v}/\sigma_{\rm cl}$), Figure \ref{fig2}) are used as proxies for mean stellar age and cluster in-fall time, respectively.
In each panel, different colors indicate different in-fall time zones separated by the vertical dashed lines and the color bars on top (Section \ref{subsec:Infall Time Estimate}): Early in-fall (red); Intermediate in-fall (orange); and Recent in-fall (light blue). \textbf{Top:} the distribution of galaxies at \textit{all redshifts} ($0.26 < z < 1.13$).
Colored diamonds indicate the bootstrapped mean \dn\ of galaxies in the corresponding in-fall time zones. The error of the mean is smaller than the symbol size.
The black solid line is a linear fit to individual galaxies derived from bootstrapping with the \dn\ uncertainties accounted for (Table \ref{tab3}).
The dotted horizontal line indicates the \dn\ $=$ 1.55 as the guideline used in the literature to separate between star-forming (\dn\ $<$ 1.55) and quiescent (\dn\ $>$ 1.55) galaxies.
The typical uncertainty of \dn\ is marked in the bottom right of each panel.
\textbf{Middle:} the distribution of \textit{low-redshift} ($0.26 < z < 0.53$) sub-samples.
\textbf{Bottom:} the distribution of \textit{high-redshift} ($0.53 < z < 1.13$) sub-samples.
Note that there is a continuous increase in the mean \dn\ of galaxies with the in-fall time proxy at all redshifts (from top to bottom panels), as also shown in the histograms on the right.
The continuous \dn\ increase with in-fall time proxy (from Recent to Early in-fall) shown in all panels suggests that galaxies become quenched as they spent a longer time in cluster environment at all redshifts $z = 0.26-$1.13 (Section \ref{subsec:entire quenching trend}).}
\label{fig3}
\end{figure*}

%;================Table 2, K-S tests between different in-fall time zones.
\begin{table*}[ht]
\centering
%\small
\begin{threeparttable}
\caption{The two-sample K$-$S test shows whether the \dn\ distributions of different in-fall time zones are statistically different (Section \ref{subsec:entire quenching trend} and the right panel of Figure \ref{fig3})\tnote{a,b,c}}
\begin{tabular}{|l|c|c|c|}
\hline \hline
 & Total-$z$ (0.26 $< z <$ 1.13) & Low-$z$ (0.26 $< z <$ 0.53) & High-$z$ (0.53 $< z <$ 1.13) \\
\hline
Early --- Intermediate & \textbf{0.002} & \textbf{0.03} & 0.3 \\
Early --- Recent & \textbf{0.00004} & \textbf{0.0007} & 0.07 \\
Intermediate --- Recent & 0.3 & 0.4 & 0.8 \\
\hline
\hline \hline
\end{tabular}
\label{tab2}
{\small
\begin{tablenotes}
\item[a] {Galaxies are classified as `Early', `Intermediate', and `Recent' in-fall galaxies based on their location in the phase space as a proxy for their in-fall stage into the cluster. (Figure \ref{fig2} and Section \ref{subsec:Infall Time Estimate}).}
\item[b] {Numbers indicate the null (i.e., false-positive) probability that a given pair of in-fall time zones' (Early, Intermediate, and Recent) \dn\ distributions are drawn from the same parent distribution.}
\item[c] {Boldfaced null probability indicates that the \dn\ distributions of a given pair of in-fall time zones are statistically different being less than 0.05.}
\end{tablenotes}
}
\end{threeparttable}
\end{table*}
%========================================================================================

%================================ Table 3 the Bootstrapped linear fit of Dn4000 vs. In-fall time proxy.===
\begin{table}%[h!]
\centering
\begin{threeparttable}
\caption{The fit relation of \dn\ vs. In-fall time proxy shows the consistent negative slope $\alpha$ across all redshift and luminosity bins (Sections \ref{subsec:entire quenching trend}--\ref{subsec:redshift and luminosity dependence}) \tnote{a}}
\begin{tabular}{lll}
\hline \hline
Redshift \tnote{b} / Luminosity\tnote{c} & Slope $\alpha$ \tnote{d} & Intercept $\beta$ \tnote{d}\\
\hline
Total-$z$  / faint$+$bright & -0.096 $\pm$ 0.016 & 1.48 $\pm$ 0.01 \\ 
Low-$z$ / faint$+$bright & -0.093 $\pm$ 0.020 & 1.54 $\pm$ 0.02 \\ 
High-$z$ / faint$+$bright & -0.078 $\pm$ 0.025 & 1.44 $\pm$ 0.02 \\ 
\hline
Total-$z$  / faint & -0.082 $\pm$ 0.022 & 1.45 $\pm$ 0.02 \\ 
Low-$z$ / faint & -0.080 $\pm$ 0.027 & 1.52 $\pm$ 0.02 \\ 
High-$z$ / faint & -0.052 $\pm$ 0.035 & 1.39 $\pm$ 0.03 \\ 
\hline
Total-$z$ / bright & -0.108 $\pm$ 0.024 & 1.53 $\pm$ 0.02 \\ 
Low-$z$ / bright & -0.096 $\pm$ 0.033 & 1.60 $\pm$ 0.04 \\ 
High-$z$ / bright & -0.095 $\pm$ 0.033 & 1.50 $\pm$ 0.03 \\ 
\hline \hline \\
\label{tab3}
\end{tabular}
{\small
\begin{tablenotes}
\item[a] The relation is described by \dn \ = $\alpha$ log [($r_{\rm proj}/r_{\rm 500}$) $\times$ ($\Delta \rm{v}/\sigma_{\rm cl}$)] $+$ $\beta$, same as Eq. \ref{Eq7}. 
\item[b] Total-$z$: 0.26 $< z <$ 1.13; Low-$z$: 0.26 $< z <$ 0.53; High-$z$: 0.53 $< z <$ 1.13.
\item[c] Galaxy $i$-band luminosity (\ilum) is used (Section \ref{subsec:Galaxy i band Luminosity}). Galaxies are divided into faint (sub-$L^{*}$, \ilum\ $ < 1$) and bright (super-$L^{*}$, \ilum\ $ > 1$) sub-samples relative to the characteristic luminosity $L^{*}$.
\item[d] The values and uncertainties are derived from 1000 bootstrapping of the fit with the \dn\ uncertainties accounted for.
\end{tablenotes}
}
\end{threeparttable}
\end{table}
%========================================================================================

We now investigate how the \dn\ (mean stellar age) of galaxies varies with time since in-fall to study environmental impacts on star formation of galaxies.
Figure \ref{fig3} shows the \dn\ versus ($r_{\rm proj}/r_{\rm 500}$) $\times$ ($\Delta \rm{v}/\sigma_{\rm cl}$) (a proxy for in-fall time, see Section \ref{subsec:Infall Time Estimate} for details).
The top panel shows the distribution of galaxies at all redshifts ($0.26 < z < 1.13$).
The colored diamond indicates the mean \dn\ of galaxies in the corresponding colored in-fall time zone, which is derived from 1000 bootstrap realizations accounting for the \dn\ uncertainties.

Notably, there is a continuous increase in the mean \dn\ of galaxies with in-fall time proxy moving from Recent to Early In-fall populations.
The increase in the mean \dn\ is from $1.51 \pm 0.01$ to $1.62 \pm 0.01$.
The net increase ($\Delta$\dn ) corresponds to an age increase of $\sim0.71 \pm 0.4$ Gyr, based on a simple stellar population modeling assuming an instantaneous burst of star formation with solar metallicity from \cite{kauf03} \footnote{The ``exact" net age increase depends on the details of assumed stellar population modeling parameters such as the shape of star formation history and metallicity (see e.g., Figure 2 of \citealt{kauf03} and Figure 1 of \citealt{hern13}). For instance, the variations in the assumed metallicity (0.4 -- 2.5 $Z_{\odot}$) result in the corresponding age increase to the \dn\ increase to vary from $\sim$ 0.31 Gyr to $\sim$ 1.12 Gyr. However, more importantly, we note that there is no qualitative change in our interpretation about the \dn\ increase as the age increase of galaxies with in-fall time proxy.}. The \textit{quantitative} age increase estimated above only describes the relative difference in mean stellar age between Recent and Early in-fall populations. This average age difference is estimated from galaxies that span a wide range of redshifts ($0.26 < z< 1.13$), and therefore the reported age difference does not capture any redshift-dependent age differences between in-fall zones.
We examine in detail the redshift dependence of age difference between in-fall zones in Section \ref{subsec:redshift and luminosity dependence}.
%----------Figure 5---------- dn4000 zoom-in with redshift and faint and bright
\begin{figure*}[ht!]
\centering
\includegraphics[width=1.\textwidth]{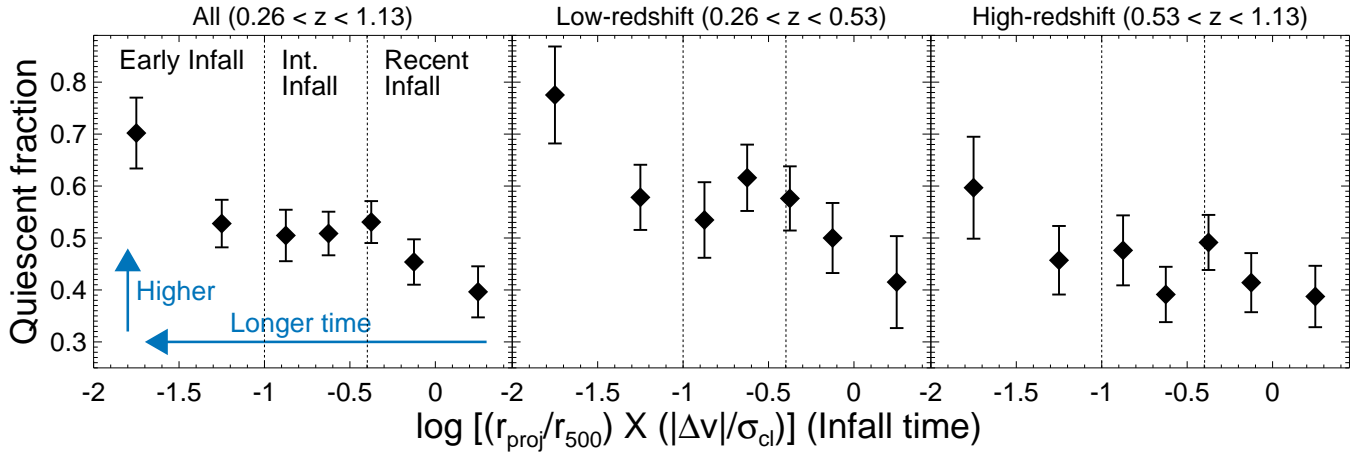}
\caption{Diagram showing that the fraction of quiescent galaxies (${f_{\rm {Q}}}$) increases with in-fall time proxy.
Each panel shows the same relationship, but for different redshift ranges as indicated on the tops of the panels.
Quiescent galaxies are defined as \dn\ $> 1.55$. 
The ${f_{\rm {Q}}}$ are calculated in bins of 0.25 or 0.5 dex (smaller bin size adopted for crowded in-fall bin) of in-fall proxy ($ {\rm log}[(r_{\rm proj}/r_{\rm 500}) \times (\Delta \rm{v}/\sigma_{\rm cl}$)).
The error bars indicate 1$\sigma$ uncertainties based on Poission statistics.
There is an increasing ${f_{\rm {Q}}}$ with in-fall proxy in all panels (i.e., regardless of redshift), although the low-$z$ galaxies (middle panel) show systematically higher ${f_{\rm {Q}}}$ compared to the high-$z$ galaxies (right panel) for any given in-fall bin (see Section \ref{subsec:redshift and luminosity dependence} for the redshift dependence of environmental quenching trends).
%Also, note a redshift dependence of quiescent fraction, such that the low-$z$ galaxies (middle panel) show higher quiescent fraction for any given in-fall bin compared to the high-$z$ counterparts (right panel) due to the redshift evolution of star-forming main sequence (Section \ref{subsec:redshift and luminosity dependence}).
The increasing quiescent fraction of galaxies with in-fall time proxy suggests that galaxies become quiescent due to longer exposure time to environmental effects.
}
\label{fig4_quie_fra}
\end{figure*}
%=========================================================================================

The continuous increase in the mean \dn\ of galaxies with in-fall proxy is also shown by the gradual shift of the \dn\ histograms of each in-fall zone in the right panel of Figure~\ref{fig3}.
That is, Early in-fall galaxies (red solid line) tend to have a larger fraction of high \dn\ (\dn \ $\gtrsim 1.5$) galaxies compared to the Recent in-fall counterparts (blue dashed line), and the Intermediate in-fall galaxies show the intermediate distribution between Recent and Early in-fall zones.
This steady shift of the distributions toward larger \dn\ strength from Recent (blue) to Early (red) in-fall populations suggests that the mean age of cluster galaxies increases with time since in-fall, although there is a broad distribution of \dn\ values in each the in-fall zone, which is reflected in the overlapping \dn\ histograms of different in-fall populations.

The different \dn\ distributions between different in-fall time zones are further supported by the two sample K-S test (Table 2).
Table \ref{tab2} presents the results of the K-S tests for the null (i.e., false-positive) probability that the two-select distributions are statistically different.
The left column of the table shows that the \dn\ distributions between Early in-fallers and both other categories of in-fallers are statistically different, with null probabilities $< 0.05$.

We also perform a linear fit to the \dn\ vs. in-fall time proxy relation for galaxies with the following form:
%========================Dn4000 vs. In-fall time proxy Linear fit Equation============================
\begin{eqnarray}
{\rm D_{n}4000} = {\alpha} \ {\rm log}[(r_{\rm proj}/r_{\rm 500}) \times (\Delta \rm{v}/\sigma_{\rm cl})] + {\beta}, 
\label{Eq7}
\end{eqnarray}
where $\alpha$ and $\beta$ are the slope and intercept of the relation, respectively.
%=========================================================================================
We perform 1000 iterations for the fits by accounting for the \dn\ uncertainties.
We find a negative slope ($\alpha$) of -0.096 $\pm$ 0.016 of the relation that is constrained to be negative at high statistical significance. This suggests that the \dn\ of galaxies increases with lower ($r_{\rm proj}/r_{\rm 500}$) $\times$ ($\Delta \rm{v}/\sigma_{\rm cl}$), meaning older age with longer time spent in clusters since in-fall.
We note however that the quantitative slope that we measure could be subject to small selection biases due to our observational selection effects, specifically that we preferentially target red sequence (passive) galaxies (see Section \ref{subsec:Sample Selection} for details).

The middle panel in Figure \ref{fig3} shows the distribution of \textit{low-redshift} ($0.26 < z < 0.53$) sub-samples.
The low-$z$ galaxies show similar trends as those of the entire sample (top panel).
That is, they also show a constant increase of \dn\ with in-fall proxy, with the fitting slope ($\alpha$) of -0.093 $\pm$ 0.020 (Table \ref{tab3}) and statistically different \dn\ distributions between Early in-fallers and other in-fallers by the K-S test (the middle panel of Table \ref{tab2}).
A noticeable difference from the entire sample is that the \dn\ of low-$z$ galaxies is on average \textit{larger} by $\sim0.05$, regardless of in-fall time proxy.
This is further shown by the larger intercept value of the fitted \dn\ vs. in-fall proxy relation in low-$z$ sub-samples compared to the entire sample in Table \ref{tab3} (i.e., 1.54 $\pm$ 0.02 vs. 1.48 $\pm$ 0.01 for low-$z$ sub-samples and the entire sample, respectively). 
The $\Delta$\dn\ of the mean \dn\ from Recent to Early in-fall is a 0.11 increase from $\sim$ 1.56, which corresponds to $\sim 0.84$ $\pm$ 0.6 Gyr age increase.

The bottom panel of Figure \ref{fig3} shows the distribution of \textit{high-redshift} ($0.53 < z < 1.13$) sub-samples.
Like the other two redshift bins (top and middle panels), the high-$z$ sub-samples also show the increasing \dn\ trends with in-fall proxy, as shown by the increasing mean \dn\ and the gradual shift of the \dn\ histogram toward larger \dn\ strength when moving from Recent to Early in-fallers.
We also found a negative slope ($\alpha$ of -0.078 $\pm$ 0.025) of the relation for high-$z$ galaxies, which is slightly shallower but still consistent with the other redshift bins within 1$\sigma$ uncertainties (Table \ref{tab3}).

The K--S test for \dn\ distributions of different in-fall zones of high-$z$ sub-samples does not show statistical significance as the related null probability that the tested distributions are drawn from the same distribution is larger than 0.05 (in the right column of Table \ref{tab2}).
Nonetheless, like the other redshift bins, the null probabilities systematically vary with in-fall time zones such that we see the same qualitative trends in \dn\ with in-fall time proxy.
These systematic K--S test results with in-fall zones at least suggest that the \dn\ distribution of our high-$z$ sub-samples is related to the galaxies' in-fall time proxy.
The $\Delta$\dn\ of the mean \dn\ from Recent to Early in-fall for high-$z$ galaxies is a $\sim 0.08$ increase from $\sim$ 1.47, which corresponds to $\sim 0.63$ $\pm$ 0.4 Gyr age increase.

Contrary to low-$z$ galaxies, the high-$z$ sub-samples show on average \textit{smaller} \dn\ values by $\sim0.05$ compared to the entire sample at all in-fall proxy.
The smaller \dn\ value of high-$z$ sub-samples is also seen by their small intercept of the \dn\ vs. in-fall proxy relation compared to the entire sample and low-$z$ sub-samples in Table \ref{tab3}.

This redshift dependence of the \dn\ strengths across in-fall proxy is likely attributed to the redshift evolution of star-forming main sequence, such that the average \dn\ strength of galaxies at fixed stellar mass decreases with increasing redshift \citep[e.g.,][]{whit12,hain17,pand17}.
We will further discuss the redshift dependence of the \dn\ distributions of our sample galaxies associated with in-fall time in Section \ref{subsec:redshift and luminosity dependence}.

The continuous galaxy age (\dn ) increase with in-fall time proxy (Figure \ref{fig3}) is also consistent with the increasing fraction of quiescent galaxies ($f_{\rm Q}$) with in-fall proxy shown in Figure \ref{fig4_quie_fra}.
In the figure, the $f_{\rm Q}$ is calculateƒ as the fraction of galaxies whose \dn\ $> 1.55$ in bins of in-fall proxy.
Indeed, the $f_{\rm Q}$ increases from $0.4 \pm 0.05$ to $0.7 \pm 0.07$ while moving from Recent to Early in-fall zones across all redshifts ($z = 0.26$--1.13).

These steady increases in \dn\ and $f_{\rm Q}$ with in-fall time proxy would suggest that galaxies that have in-fallen into clusters earlier are quenched earlier due to longer exposure time to environmental effect.
Our findings are qualitatively consistent with previous studies for several high-$z$ ($z \sim1$) clusters \citep{nobl13,nobl16,wern22} and local ($z < 0.2$) clusters \citep{pasq19,smit19,upad21}, as well as recent cluster simulations showing systematic suppression of star formation of galaxies since in-fall \citep{wetz13,oman16,rhee20,oman21,coen21}.
Our time-averaged kinematic analysis spans a wide range of redshift, $z=0.26$--1.13, with a sample of clusters that are uniformly selected above an approximately constant mass threshold\footnote{We note that our results of increasing \dn\ with in-fall time proxy virtually do not change even when the sample clusters are divided into specific cluster mass criteria (e.g., the expected mass-growth curve with redshift \citealt{fakh10}) within the sample clusters' mass range $3 \times 10^{14} M_{\odot} \lesssim M_{\rm 500} \lesssim 3 \times 10^{15} M_{\odot}$ (Section \ref{subsec:Clusters from SPT and ACT})}, and reveals a remarkably consistent picture of environmental quenching extending out to $z \sim1$.

For instance, environmental mechanisms such as ram pressure stripping \citep{gunn72,abad99,ebel14}, starvation \citep{lars80}, harassment \citep{moor96}, and tidal interactions \citep{byrd90} have been suggested as the ones enabling to suppress the star formation of galaxies in clusters.
Indeed, a systematic gas stripping process has been observed in the Virgo and Abell 963 clusters based on the projected phase space analysis \citep{voll01,bose14b,jaff15,yoon17,moro21}, where the level of {\HI} gas stripping is closely related to the in-falling stages identified in the phase space (see e.g., Figure 5 of \citealt{yoon17}).
Furthermore, the stripping of other components of galaxies, such as warm dust \citep{nobl16}, molecular gas (CO) \citep{fuma09,bose14a,lee18}, and dust \citep{cort16,long20}, are also reported in cluster environments.
These results clearly show that gas stripping is actively occurring to cluster galaxies (likely) through the ram pressure by the hot intracluster medium (ICM) in the deep gravitational potential well of clusters.
This is further supported by the recent cosmological cluster simulation that shows the systematic gas depletion of galaxies by ram pressure stripping since in-fall \citep{jung18}.

It is not unreasonable to expect that such ram pressure stripping has occurred to our sample cluster galaxies as well, considering the similar mass range (thus gravitational potential well) of our sample clusters with those reported for gas stripping (i.e., 1--8 $\times 10^{14} M_{\odot}$).
Indeed, the significant X-ray detection (X-ray luminosity $> 10^{44} \ \rm{erg} \ \rm{s}^{-1}$) and the associated hot ICM temperature ($>$ 2 $\times 10^{7}$ K) are found in our cluster sub-samples across all redshifts \citep{mcdo14,bulb19}.
However, ram-pressure stripping should be most effective on low-mass systems \citep[e.g.,][]{janz2021}, and the luminosity cut of our analysis (i.e., \ilum \ $> 0.35$, Section \ref{sec:Sample Selection and Data Analysis}) restricts our sample to moderately massive galaxies, suggesting that we should not be highly sensitive to the effects of the ram-pressure mechanism.

A slow quenching process, such as the gradual shut off of gas supply \citep[``starvation''; see, for example][]{lars80} seems a plausible quenching process driving the gradual age (\dn ) trend with in-fall proxy in our sample galaxies (Figure \ref{fig3}). Physically, this process is one that acts via the truncation of the galaxy's circumgalactic medium (CGM), rather than a faster mechanism that achieves the rapid removal of ISM gas. A slow starvation quenching process would take effect over the course of several dynamical crossing times after a galaxy has fallen into the galaxy cluster environment, which is in good agreement with our measurement of a small average net age increase from Recent in-fall to Early in-fall populations (i.e., $0.71 \pm 0.4$ Gyr) compared to the typical dynamical time scales of $1-3$ Gyr of cluster galaxies \citep[e.g.,][]{rhee20}.

A slow quenching trend is also consistent with the approximately constant slope ($\alpha$) that we measure for the mean stellar population age versus in-fall proxy relation (i.e., the \dn \ vs. $\alpha$ log [($r_{\rm proj}/r_{\rm 500}$) $\times$ ($\Delta \rm{v}/\sigma_{\rm cl}$)] $+$ $\beta$ relation (Eq. \ref{Eq7})).
Specifically, the slope $\alpha$ does not significantly depend on galaxies' \ilum \ (stellar mass), as both ``bright'' (\ilum $> 1$) and and ``faint'' (\ilum $< 1$) sub-samples have statistically similar $\alpha$ ($\sim -0.09$) at all redshifts (Table \ref{tab3}). The similar $\alpha$ we measure across galaxy luminosity suggest that the environmental quenching mechanism acting on these galaxies is not strongly dependent on either galaxy stellar mass or orbital velocity, which matches the expectation for the starvation effect in cluster environments. Qualitatively, our results are in broad agreement with the known ``delayed-then-rapid'' quenching processes \citep[e.g.,][]{wetz13,hain15,gall21}.

We also note that while the average net age increase ($0.7 \pm 0.4$ Gyr) seen in our sample galaxies appears to be broadly consistent with the quenching age ($0.1-1.3$ Gyr) by the ram pressure gas stripping in local galaxy clusters \citep{crow08,bose16}, it should be cautious to \textit{directly} compare the age increase we measure with the quenching ages from the literature. 
This is due to the difference in how the quenching-related ages are measured between the studies such that our measurement is based on the average net age increase of stellar populations since infall, whereas the SFR quenching age often measured in the literature is considered the time taken for a galaxy to transform from gas-rich, star-forming, to a totally quenched passive system.

We also note that there are other environmental mechanisms that are likely at play in our sample galaxies.
Of them, the preprocessing---i.e., pre-exposure of group-scale environmental effects before in-falling to the main cluster \citep{balo00,fuji04,han18,lee22}---is known to be able to quench the star formation of in-falling galaxies even in the cluster outskirts ($\gtrsim R_{\rm vir}$) \citep{balo00}, the region where cluster environmental effects are expected to be insignificant.
Indeed, recent observations for high redshift clusters ($z \sim 1$) suggest that the majority of massive galaxies are quenched during in-fall in the cluster outskirts ($1< R/R_{\rm 200} < 3$), which might imply the pre-processing effects prior to in-fall \citep{wern22}.

While our analysis alone cannot pin down the most likely environmental mechanisms responsible for the quenching of cluster galaxies, we emphasize that the steady \dn\ increase with in-fall proxy in Figure \ref{fig3} strongly indicates the environmental quenching at play over a wide span of redshift (0.26 $< z <$ 1.13).
Also, our results are qualitatively consistent with the recent work at similar redshifts ($0.3 < z <1.5$) \citep{webb20,khul21}, where the authors used stellar population fitting analysis to measure star formation histories and ages of quiescent cluster galaxies.

%-----------Section 4.2-----------------------------
\subsection{Luminosity Dependence of Environmental Quenching: Trends with Faint (\ilum\ $< 1$) and Bright (\ilum\ $> 1$) Galaxies}
\label{subsec:luminosity dependence on the quenching}
In the previous section, we have shown that there is a continuous increase of \dn\ with in-fall time proxy, which suggests that galaxies that have in-fallen earlier are quenched earlier due to environmental effects. In this section, we investigate a galaxy luminosity dependence of the environmental quenching trend.
We divide our sample galaxies into `faint' (sub-$L^{*}$, \ilum $< 1$) and `bright' (super-$L^{*}$, \ilum $> 1$) sub-samples based on their $i$-band luminosity (Section \ref{subsec:Galaxy i band Luminosity}).

%----------Figure 4----------Mass quenching with environmental quenching for all redshift (faint and bright bins)
\begin{figure}[ht!]
\centering
\includegraphics[width=0.48\textwidth]{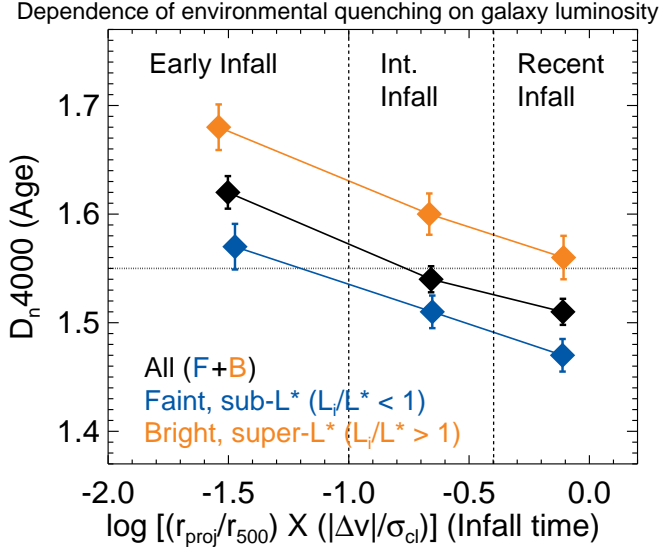}
\caption{
Similar to the top panel of Figure \ref{fig3}, but focusing on the galaxy luminosity dependence of the environmental quenching effect (i.e., the continuous age (\dn ) increase of galaxies since in-fall).
The diamonds and the associated error bars are the bootstrapped mean \dn\ strength and the 1$\sigma$ standard deviation of galaxies in each in-fall zone.
Different colors indicate the different $i$-band luminosity bins: All (black), faint (sub-$L^{*}$, orange), and bright (super-$L^{*}$, blue).
Note that both faint and bright sub-samples show a continuous \dn\ increase with in-fall proxy. 
However, at all fixed in-fall zones, bright galaxies have larger mean \dn\ compared to the faint counterparts.
This suggests that environmental quenching impacts galaxies regardless of galaxies' luminosity, yet the exact age (\dn ) depends on the galaxies' luminosity due to internal mass quenching (i.e., downsizing) effect at fixed environment (Section \ref{subsec:luminosity dependence on the quenching}).}
\label{fig4}
\end{figure}
%----------------------------------------------------------------------------

%----------Figure 4----------Mass quenching with environmental quenching for all redshift (faint and bright bins)
\begin{figure*}[ht!]
\centering
\includegraphics[width=0.8\textwidth]{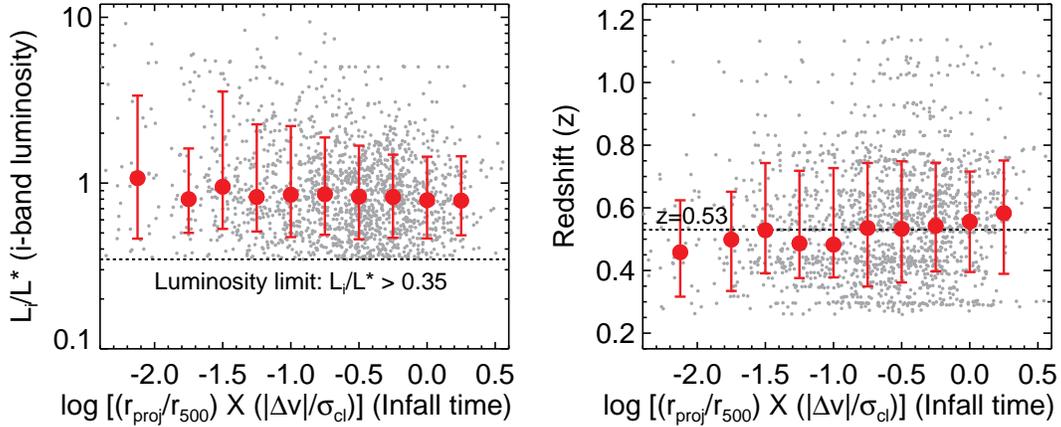}
\caption{The figure shows that galaxy luminosity (left) and redshift (right) do not show an obvious dependence on in-fall proxy.
The red points with error bars indicate the median values and the associated 1$\sigma$ spread per in-fall proxy bin.
\textbf{Left:} the galaxy luminosity (\ilum) distribution is similar regardless of in-fall proxy, with median values consistent within 1$\sigma$.
\textbf{Right:} the redshift distribution is also consistent across all in-fall proxy, although the median redshift seems to slightly decrease with in-fall proxy within 1$\sigma$ population spread.
Little (or no) dependence of luminosity and redshift on in-fall proxy suggests that the increasing \dn\ trend with in-fall time proxy (e.g., Figure \ref{fig3} and Section \ref{subsec:entire quenching trend}) is not driven by either luminosity or redshift. 
Rather, it is most likely due to the environmental quenching such that galaxies that have in-fallen earlier stopped star formation earlier.
And it is \textit{regardless of} specific galaxy luminosity (Figures \ref{fig4} and \ref{fig4_5_eps2} and Section \ref{subsec:luminosity dependence on the quenching}) and redshift bins (Figure \ref{fig5} and Section \ref{subsec:redshift and luminosity dependence}).}
\label{fig4_i_z_infall}
\end{figure*}
%----------------------------------------------------------------------------

%----------Figure 4_5----------In-fall vs. Luminosity with Dn4000 color-coded
\begin{figure*}[ht!]
\centering
\includegraphics[width=0.6\textwidth]{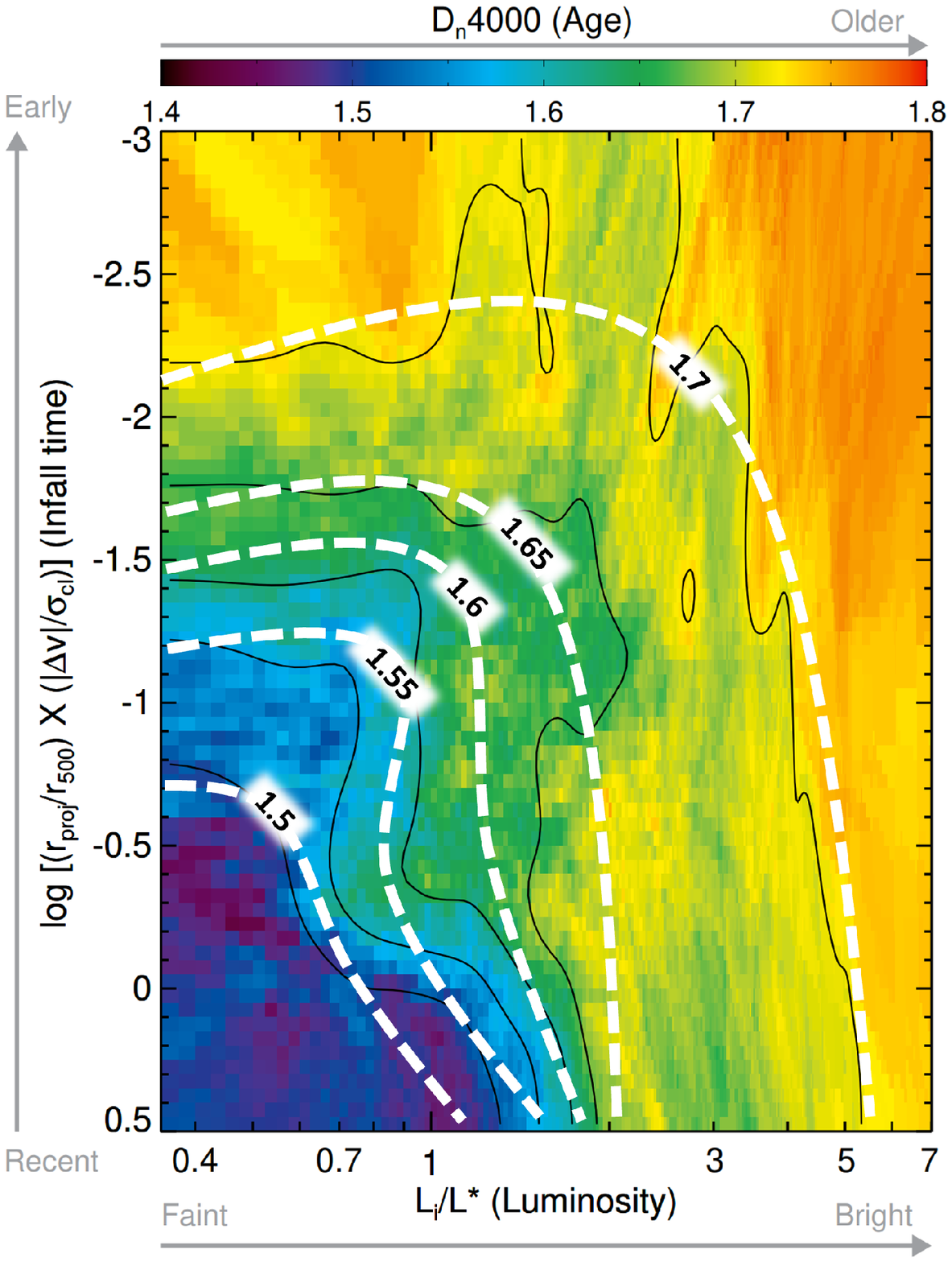}
\caption{Diagram showing that environment and luminosity are the two main drivers of galaxy quenching where the environmental effect is \textit{kinematically} measured from the phase space of cluster galaxies (Figure \ref{fig2} and Section \ref{subsec:Infall Time Estimate}).
The color indicates the mean \dn\ strength of the nearest 100 sample galaxies at a given luminosity and in-fall proxy location with a grid spacing of 0.025 \ilum\ and 0.05 ${\rm log}[(r_{\rm proj}/r_{\rm 500}) \times (\Delta \rm{v}/\sigma_{\rm cl})]$.
The white dashed lines are the contours of several \dn\ values tracing the mean \dn\ distributions of sample galaxies.
It is clear that the mean \dn\ typically increases with in-fall proxy for any luminosity, which suggests that environmental quenching affects galaxies regardless of luminosity (Section \ref{subsec:luminosity dependence on the quenching}).}
\label{fig4_5_eps2}
\end{figure*}
%----------------------------------------------------------------------------

Figure \ref{fig4} shows the same \dn\ versus in-fall time proxy diagram as in the top panel of Figure \ref{fig3}, but focusing on the galaxy luminosity dependence by splitting into faint and bright sub-samples.
The black diamonds and the associated error bars are the bootstrapped mean \dn\ strength and the 1$\sigma$ spread of the full sample, same as in the top panel of Figure \ref{fig3}.
The orange and blue colors indicate the same symbols as black, but for faint and bright sub-samples, respectively.

Notably, both faint and bright sub-samples show qualitatively the same environmental quenching trend with in-fall time, as both sub-populations show the continuous increase of \dn\ with the in-fall time proxy.
The faint and bright sub-samples also have slopes ($\alpha$) of \dn\ vs. in-fall proxy relation consistent within 1 $\sigma$, with $\alpha= -0.082 \pm 0.022$ and $-0.108 \pm 0.024$ for faint and bright galaxies, respectively as in Table \ref{tab3}).
This means that the environmental quenching affects galaxies regardless of galaxy luminosity (stellar mass)\footnote{Note that while the $i$--band luminosity (\ilum) is a useful quantity for optical light of stellar populations, it is only a rough proxy for stellar mass and is subject to significant mass-to-light (\textit{M/L}) ratio variations.
This caveat especially applies for galaxies at high redshift ($z > 0.8$) where the $i$--band samples rest-frame light blueward of the 4000 $\rm{\AA}$ break.
Only a small fraction ($8 \%$) of our galaxies are at z $> 0.8$, so that the large majority of our analysis is based on galaxies where the $i$--band samples rest-frame light redward of the 4000 $\rm{\AA}$ break.}.

However, there is a noticeable difference between faint and bright sub-samples, such that bright sub-samples always have larger mean \dn\ strength than the faint counterparts across in-fall zones.
This trend does not arise from any potential dependence of luminosity on in-fall proxy in the sense that bright galaxies might be preferentially located towards earlier in-fall zones and thus have larger mean \dn\ than the faint counterparts. 
We show that the luminosity distribution of sample galaxies is similar across all in-fall proxy in the left panel of Figure \ref{fig4_i_z_infall}.

Thus, at \textit{fixed} in-fall zone (environmental effect), this quantitative \dn\ difference between faint and bright sub-samples is likely attributed to the mass-dependent ``internal'' quenching of galaxies \citep[e.g.,][and references therein]{gava96,bose01,peng10,kim16,kim18}. That is, more massive (bright) galaxies become quenched earlier than the less massive (faint) counterparts.
This mass quenching may be attributed to mechanisms such as virial shock heating of in-falling gas in massive galaxy halos \citep[e.g.,][]{deke06} and/or secular AGN feedback from the supermassive black holes in galaxy center \citep[e.g.,][]{choi15,bluc22}.

The individual effects of environment-related in-fall time proxy and galaxy luminosity (\ilum ) on quenching as traced by mean galaxy age are demonstrated in Figure \ref{fig4_5_eps2}.
Figure \ref{fig4_5_eps2} is generated by creating a density map of the spatially averaged \dn\ value within a grid of in-fall proxy and luminosity. Each pixelated grid position is 0.025 in \ilum\ by 0.05 in ${\rm log}[(r_{\rm proj}/r_{\rm 500}) \times (\Delta \rm{v}/\sigma_{\rm cl})]$, and within eaech grid pixel we measure the mean \dn\ strength of the nearest 100 sample galaxies. Due to some regions of the grid space being very sparsely sampled we then apply a 2-dimensional Gaussian smoothing kernel with a width of 3 grid pixels to improve the visualization of the resulting density map.
Clearly, the mean \dn\ (age) of galaxies is a function of \textit{both} environment and luminosity as the colored \dn\ strength continuously increases with in-fall proxy (y-axis) and luminosity (x-axis).
However, when fixing one parameter (i.e., x or y-axis), the mean \dn\ strength increases along with the other parameter.
This means that a continuous \dn\ increase along with y-axis (in-fall proxy) is due to the environmental quenching effect at any given luminosity. 
This trend is also consistent with the `knee'-shaped white dashed lines that are the contours of constant \dn\ strengths tracing the mean \dn\ distributions of sample galaxies.

Therefore, Figures \ref{fig4} and \ref{fig4_5_eps2} suggest that environmental quenching impacts galaxies independent of galaxy luminosity, although the exact age (\dn ) depends on the galaxies' luminosity due to the internal mass quenching mechanism. 
Our kinematic environmental results through in-fall time proxy show a consistent picture where stellar mass and environment are the two main drivers of galaxy quenching \citep[e.g.,][]{peng10,smit12,kawi17,sobr21}.

%----------Figure 5---------- dn4000 zoom-in with redshift and faint and bright
\begin{figure*}[ht!]
\centering
\includegraphics[width=1.\textwidth]{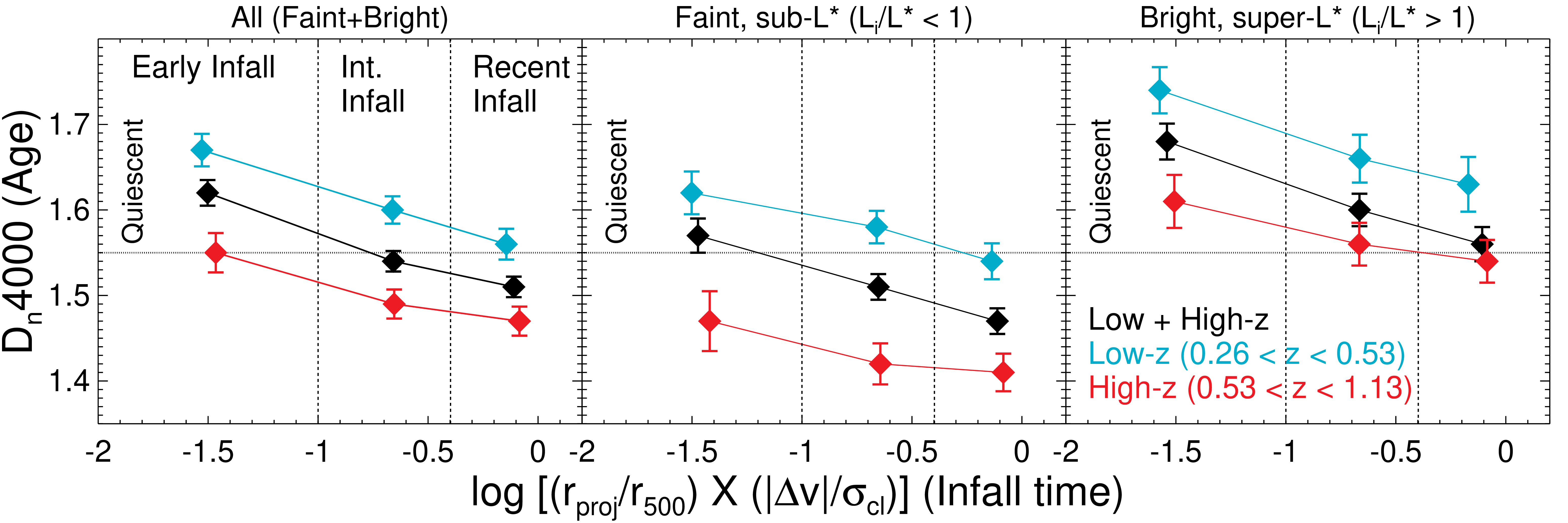}
\caption{
Diagram showing the trends of quenching since in-fall (i.e., a continuous increase of \dn\ with in-fall time proxy) across redshift and luminosity bins.
In all panels, different colors indicate different redshift bins; \textbf{black:} all-$z$ (0.26 $< z <$ 1.13), \textbf{blue:} low-$z$ (0.26 $< z <$ 0.53), and \textbf{red:} high-$z$ (0.53 $< z <$ 1.13).
Different panels show different luminosity bins; \textbf{left:} All (faint+bright), \textbf{middle:} faint (sub-$L^{*}, $\ilum\ $< 1$), and \textbf{right:} bright (super-$L^{*}, $\ilum\ $> 1$).
The diamonds with error bars indicate the bootstrapped mean \dn\ and the associated standard deviation of sample galaxies in each in-fall zone, which also accounts for \dn\ measurement uncertainties.
The formats are the same as in Figure \ref{fig3}.
By comparing the blue and red points in \textit{each} panel, it is clearly shown that low-$z$ galaxies have larger mean \dn\ than the high-$z$ counterparts at fixed in-fall zone (fixed environmental effect), regardless of luminosity.
This is due to the redshift dependence of star-forming main sequence (Section \ref{subsec:redshift and luminosity dependence}).
Also, by comparing the middle and right panels for the \textit{same} colored points, it is shown that bright galaxies have larger mean \dn\ than the faint counterparts at fixed in-fall zone across redshift, which is due to the mass-dependent quenching effect (Section \ref{subsec:luminosity dependence on the quenching}).
Thus, by dividing sample galaxies into specific redshift and luminosity bins, this figure clearly shows that environmental quenching \textit{does} exist and impact galaxies since in-fall into clusters, at all redshift and luminosity, yet the exact age (\dn ) depends on both redshift and luminosity at fixed environment (in-fall zone) (Section \ref{subsec:redshift and luminosity dependence}).
}
\label{fig5}
\end{figure*}

%-----------------Section 4.3--------------------------------------------------
\subsection{Environmental Quenching Observed Across Wide Ranges of Redshift (0.26 $< z <$ 1.13) and Luminosity (0.35 $<$ \ilum\ $<$ 10)}
\label{subsec:redshift and luminosity dependence}
So far, we have demonstrated that environmental quenching occurs since in-fall such that galaxies' \dn\ continuously increases as moving from Recent to Early in-fallers (Figure \ref{fig3}).
Also, the environmental quenching seems to impact galaxies regardless of their luminosity (Figures \ref{fig4} and \ref{fig4_5_eps2}).
We now control for redshift and luminosity \textit{simultaneously} to further isolate environmental quenching effects, as both parameters can independently alter the \dn\ strengths (similarly sSFR) of galaxies, regardless of environments \citep[e.g.,][]{kauf03,hain17,kim18}.

Figure \ref{fig5} shows the same \dn\ trends with in-fall time proxy as in other figures, but further divided by specific redshift and luminosity bins.

First, in all panels showing different luminosity bins, low-$z$ galaxies (blue points) clearly show larger mean \dn\ values than the high-$z$ counterparts (red points) across in-fall zones.
At the same time, both redshift bins show a continuous increase of mean \dn\ with in-fall time proxy due to environmental quenching at play at all redshift bins.
This systematic redshift dependence of \dn\ at fixed in-fall proxy (environmental effect) and luminosity is likely attributed to the redshift-dependent star-forming main sequence (SFMS) of galaxies \citep[e.g.,][]{whit12,sobr14,pand17}.
As the specific star formation rates of galaxies for a given stellar mass decrease with decreasing redshift, the \dn\ strength of galaxies given luminosity is also expected to \textit{increase} with decreasing redshift.
Note that this trend does not arise from any potential dependence of galaxy redshift on in-fall proxy in the sense that low-$z$ galaxies might be preferentially located towards earlier in-fall zones and thus have larger mean \dn\ than the high-$z$ counterparts. 
This is shown in the right panel of Figure \ref{fig4_i_z_infall} where the redshift distribution of sample galaxies is mostly similar across all in-fall proxy, although the median redshift appears to slightly increase with in-fall proxy within 1$\sigma$ population spread (i.e., $\Delta z \sim 0.1$ across in-fall proxy).

Also, by fixing redshift bins (i.e., focusing on the same colored points across panels), it is clear that bright (super-$L^{*}$) galaxies have larger mean \dn\ values than the faint (sub-$L^{*}$) galaxies at all in-fall zones, while both luminosity sub-samples show an environmental quenching effect (a continuous increase of \dn\ with in-fall time proxy).
As noted in Section \ref{subsec:luminosity dependence on the quenching}, this luminosity dependence of \dn\ is likely explained by the mass quenching effect in that more massive (bright) galaxies tend to stop star formation earlier than those with less mass (faint) \citep[e.g.,][]{peng10,hain17,kim16,kim18}.
By splitting into low and high-redshift bins, we further show that this luminosity dependence of \dn\ at fixed in-fall proxy exists independent of specific redshift bins.

The trends shown in Figure \ref{fig5} further suggests that environmental quenching indeed affects galaxies since in-fall regardless of redshift \textit{and} luminosity, as clearly shown by the systematic increase of the mean \dn\ with in-fall time proxy across all redshift and luminosity bins.
This is also supported by the similar gradient (slope $\alpha$) of \dn\ vs. in-fall proxy relation for all sub-samples (Table \ref{tab3}) that are consistent within uncertainties.
However, it is also evident that the exact \dn\ strength (age) at \textit{fixed} environmental effect shows the systematic dependency on redshift and luminosity due to their individual effects on \dn\ as discussed above.

%-----------Section 4.4-----------------------------
\subsection{Caveats on Projected Phase Space}
\label{subsec:projection effects}
Despite the exceptional usefulness of the projected phase space to investigate environmental effects, it is also worth noting that there is a spread in our in-fall time proxy due to sources of uncertainty such as projection effects and variations in the orbital parameters (e.g., velocity dispersion anisotropy profiles, see \citealt{capa19}).
These include a difference between the projected 2D clustercentric distance and the actual 3D distance and a mix of in-falling and out-falling (e.g., back-splash and mere interlopers) populations in the projected phase space.

Cluster simulations \citep{gill05,rhee17} have demonstrated that the line-of-sight velocity distribution of back-splash galaxies, which have passed the first pericenter and are currently outbound, overlap with the in-falling galaxies at low velocity regime ($\lesssim 400$ km/s) in the cluster outskirt (1--2 $R_{\rm 200}$) \citep[see i.e., Figure 8 of][]{gill05}. These back-splash populations thus are likely mixed with the in-falling galaxies in the projected phase space.

As a result, such sources of uncertainty in the projected phase space possibly cause a moderate spread ($\gtrsim 1.5$ Gyr) in the mean in-fall time of each in-fall zone, based on the comparison with the cluster simulation at $z=0$ \citep[see i.e., Table 1 and Figure A1 of][]{pasq19,smit19,rhee17,rhee20}.
Nevertheless, such comparison clearly shows that the \textit{mean} in-fall time of each in-fall zone continuously increases from Recent ($\sim 2.7$ Gyr) to Early ($\sim 5$ Gyr) in-fallers.
This suggests that 1) the projected phase space is indeed a powerful tool to study the average properties of cluster galaxies as a function of in-fall time, and 2) projected radial studies can only be \textit{more} affected by these projection effects than phase space studies.
More importantly, we note that these sources of uncertainty only make the trends in Figures \ref{fig3}, \ref{fig4}, \ref{fig4_5_eps2}, and \ref{fig5} shallower, which means that the trends would rather be more significant if there were no such sources of uncertainty.

%=======================Summary and Conclusions=======================
\section{Summary and Conclusions}
\label{sec:Summary and conclusions}
We have investigated the environmental quenching effects using a large sample of clusters over a wide span of redshift.
A uniform set of clusters and spectroscopically confirmed member galaxies enable us to study the quenching effects \textit{kinematically} by measuring the average time spent in cluster environments spanning a huge range in cosmic time of $\sim$ 5.2 Gyr, up to $z \sim 1$ ($0.26< z < 1.13$).
For that, we have mapped the location of galaxies in the cluster phase-space diagram to their mean in-fall time.

We found that the age-sensitive \dn\ strength of galaxies continuously increases with in-fall time proxy.
This means that galaxies that spent a longer time in the cluster environment are quenched earlier, likely due to longer exposure time to environmental effects such ram-pressure stripping and strangulation.
The most notable findings are summarized as follows:
\begin{itemize}
\item[1.] We utilize the projected phase space (i.e., clustercentric radius ($r_{\rm proj}/r_{\rm 500}$) vs. normalized line-of-sight velocity ($\Delta \rm{v}/\sigma_{\rm cl}$)) of cluster galaxies over a large redshift baseline ($z = 0.26$--1.13) to estimate the galaxies' kinematic in-fall stages (Figures \ref{fig2_sch} and \ref{fig2}).
For that, we sample 105 clusters (median mass of $4.97 \times 10^{14} M_{\odot}$) and 1626 cluster galaxies from the SPT-SZ and ACT-SZ cluster surveys and the optical follow-ups (Figure \ref{fig1} and Table \ref{tab1}).

\item[2.] We classify galaxies by their in-fall time proxy. 
Specifically, we use the caustic profiles of the phase space (i.e., the lines of constant ($r_{\rm proj}/r_{\rm 500}$) $\times$ ($|\Delta \rm{v}|/\sigma_{\rm cl}$)) as in-fall time proxy to split galaxies into ``Early'', ``Intermediate'', and ``Recent'' in-fallers (Section \ref{subsec:Infall Time Estimate}).
The Early and Intermediate in-fallers are found to be gravitationally bound by the potential well of clusters, while Recent in-fallers are either gravitationally bound or approaching the boundary, based on the comparison with the escape velocity radial profile of typical clusters (Figure \ref{fig2}).

\item[3.] Notably, a continuous \dn\ (proxy for stellar age) increase with in-fall proxy (from Recent to Early in-fall) is found in galaxies at all redshifts, $z = 0.26$--1.13 (Figure \ref{fig3}).
The increase in the mean \dn\ is statistically significant showing the increase from $1.51 \pm 0.01$ to $1.62 \pm 0.01$.
This corresponds that Early in-fall populations are, on average, $\sim 0.71 \pm 0.4$ Gyr older than the Recent in-fall counterparts, based on the net \dn\ increase ($\Delta$\dn ) of the mean \dn\ strengths of Recent to Early in-fall populations and a simple stellar population modeling (Section \ref{subsec:entire quenching trend}).
This trend is consistent with a higher fraction of quiescent galaxies having smaller in-fall time proxy values, i.e., having spent more time in the cluster environments (Figure \ref{fig4_quie_fra}).
This suggests that galaxies that spent a longer time in hosts are quenched earlier, likely due to longer exposure time to environmental effects (e.g., ram-pressure stripping and strangulation).
Compared to the typical dynamical times scales of $1-3$ Gyr of cluster galaxies \citep[e.g.,][]{rhee20}, the quenching age in our sample galaxies is small (i.e., $ 0.71 \pm 0.4$ Gyr), which is qualitatively consistent with a slow quenching starvation process \citep[e.g.,][]{lars80}  and/or the known ``delayed-then-rapid'' quenching process \citep[e.g.,][]{wetz13,hain15,gall21}.

\item[4.] The quenching trend with in-fall proxy is found \textit{regardless of} specific galaxy luminosity and redshift bins.
Specifically, both faint (sub-$L^{*}$) and bright (super-$L^{*}$) sub-samples show the continuous increase of mean \dn\ strengths with in-fall proxy (Figure \ref{fig4}).
This still holds when the sub-samples are further divided into low-$z$ and high-$z$ bins (Figure \ref{fig5}), and is also supported by the similar gradient (slope $\alpha$) of \dn\ vs. in-fall proxy relation across all redshift and luminosity bins (Table \ref{tab3}).
However, the \textit{exact} \dn\ strengths at fixed in-fall proxy (fixed environmental effect) depend on galaxy luminosity and redshift. 
That is, bright galaxies show larger \dn\ than faint galaxies for any redshift bins, and low-$z$ galaxies show larger \dn\ than the high-$z$ counterparts for any luminosity bins. 
This indicates that while galaxies experience environmental quenching since in-fall, their \textit{absolute} mean age at fixed environmental effect depends on internal mass (luminosity) quenching (Figure \ref{fig4_5_eps2} and Section \ref{subsec:luminosity dependence on the quenching}) and the redshift evolution of star-forming main sequence (Section \ref{subsec:redshift and luminosity dependence}).
\end{itemize}

This work substantially extends previous findings and provides crucial evidence for the environmental quenching effects at play outside the local Universe.
Our findings are uniquely achieved by using both a wide range of redshift and the time-averaged kinematic approach in a uniformly selected sample of clusters with nearly flat mass sensitivity.
Additionally, our results demonstrate a continuous \dn\ increase of galaxies with in-fall time proxy, in good agreement with cluster simulations that show the systematic gas stripping and quenching of galaxies since in-fall.
Future studies on the gas properties of these high-$z$ cluster galaxies combined with the phase space analysis will provide additional insights on environmental effects at high redshifts.

%====================Acknowledgement
\vspace{5mm}
We thank the referee for constructive comments that improved the quality of the paper.
This work was performed in the context of the SouthPole Telescope scientific program. SPT is supported by the National Science Foundation through grants OPP-1852617. Partial support is also provided by the Kavli Institute of Cosmological Physics at the University of Chicago. PISCO observations are supported by NSF AST-1814719. Argonne National Laboratory's work was supported by the U.S. Department of Energy, Office of High Energy Physics, under contract DE-AC02-06CH11357. 
G.M. acknowledges funding from the European Union's Horizon 2020 research and innovation programme under the Marie Skłodowska-Curie grant agreement No MARACAS - DLV-896778.
AS is supported by the ERC-StG ‘ClustersXCosmo’ grant agreement 716762, by the FARE-MIUR grant 'ClustersXEuclid' R165SBKTMA, and by INFN InDark Grant.
The Melbourne group acknowledge support from the Australian Research Council's Discovery Projects scheme (DP200101068).


%==============================References===========================================================

\clearpage

%\vspace{5mm}
%\facilities{HST(STIS), Swift(XRT and UVOT), AAVSO, CTIO:1.3m,


%\software{astropy \citep{2013A&A...558A..33A,2018AJ....156..123A},  
%          Cloudy \citep{2013RMxAA..49..137F}, 
%          Source Extractor \citep{1996A&AS..117..393B}
%          }

\end{document}